\documentclass[a4paper,12pt]{article}

\usepackage[numbers]{natbib}
\usepackage[a4paper,margin=2.5cm]{geometry}
\usepackage{tocloft}
\usepackage{caption}
\usepackage{authblk}
\usepackage{subcaption}
\usepackage{amssymb}
\usepackage{amsmath,bm}
\usepackage{layout}
\usepackage{soul}
\usepackage{color}
\usepackage{braket}
\usepackage{amsmath} 
\usepackage{hyperref}
\usepackage{dirtytalk}
\usepackage{mathrsfs}
\usepackage{xcolor}
\usepackage{graphicx}
\usepackage{listings}

\usepackage{xcolor}
\usepackage{amsmath,amssymb,amsfonts}
\usepackage{bm} 
\usepackage{algorithmic}
\usepackage{tabularx}
\usepackage{hyperref}
\hypersetup{
  colorlinks=true,
  citecolor=blue,
  linkcolor=blue,
  urlcolor=blue
}
\usepackage{array}
\usepackage{tabularx}
\usepackage{booktabs}
\usepackage{soul}
\usepackage{siunitx}

\title{End-to-End Quantum Machine Learning for Thermoelectric Materials: Projected Features, Anchored Classifiers, and Real-Hardware Demonstration}

\author[1,2]{Pietro Asinari
\thanks{Corresponding author: pietro.asinari@polito.it}}

\author[3]{Matteo Maria Piredda}

\author[1]{Giulio Barletta}

\author[1]{Matteo Fasano}

\author[3]{Nikolaos Prasianakis}

\author[1,2]{Eliodoro Chiavazzo}

\affil[1]{{\footnotesize Dipartimento Energia, Politecnico di Torino, Corso Duca degli Abruzzi 24, 10129, Torino (TO), Italy}}

\affil[2]{{\footnotesize Istituto Nazionale di Ricerca Metrologica, Strada delle Cacce 91, 10135, Torino (TO), Italy}}

\affil[3]{{\footnotesize Paul Scherrer Institute, Forschungsstrasse 111, 5232, Villigen, Switzerland}}

\date{arXiv version}

\begin{document}
\maketitle

\hrule
\vspace{0.5em}
\begin{abstract}
Quantum machine learning offers potentially rich feature spaces and nonlinear transformations, but its practical value for scientific classification remains application dependent. This work investigates the identification of thermoelectric materials from a database containing 10,797 samples and 145 descriptors, together with a 10-feature representation obtained through feature selection. Beyond the standard variational quantum classifier, we introduce two hybrid strategies. Projected Quantum Features directly use quantum expectation values as transformed inputs to a classical classifier, whereas the original Logistic-Regression-Anchored Variational Quantum Classifier trains a quantum correction around a logistic-regression baseline, thereby exploiting quantum nonlinearities without discarding an effective classical decision boundary. We place all three models within a unified formalism based on assigning trainable parameters to the quantum state, the observable, or both. End-to-end numerical experiments compare the proposed models with classical baselines under identical material-wise training, validation, and held-out testing partitions. Projected Quantum Features provide the strongest overall quantum performance, while the anchored classifier frequently outperforms the unconstrained variational classifier; nevertheless, XGBoost remains highly competitive, highlighting the challenge of establishing a practical quantum advantage on classical materials data. Finally, selected models are trained and tested on the five-qubit IQM Lagrange quantum processor using finite-shot measurements. To the best of our knowledge, this is the first thermoelectric-materials classification study in which both quantum-model training and prediction on unseen samples are performed entirely on real quantum hardware.
\end{abstract}

\vspace{0.5em}
\hrule

\vspace{0.5cm}

%\newpage
%\tableofcontents
%\newpage

%% Main text
%%%%%%%%%%%%%%%%%%%%%%%%%%%%%%%%%%%%%%%%%%%%%%%%%%%%%%%%%%%%
%%%%%%%%%%%%%%%%%%%%%%%%%%%%%%%%%%%%%%%%%%%%%%%%%%%%%%%%%%%%

\section{Introduction}
\label{sec:introduction}

Artificial Intelligence (AI) in materials science is driving significant advancements in the discovery of advanced materials for energy applications. 
The transition toward a sustainable and secure energy system critically depends on the availability of advanced materials, from thermoelectric converters and photovoltaic absorbers to electrode and electrolyte materials for batteries.
Traditionally, the discovery of such materials has relied on a combination of chemical intuition, experimental trial and error, and first-principles calculations, an approach whose cost and timescales are poorly matched to the size of the chemical space to be explored. 
AI, and machine learning (ML) in particular, is increasingly reshaping this landscape~\cite{butler2018}. 
Data-driven models can learn structure–property relationships from existing computational and experimental data, act as fast surrogates of expensive simulations, and prioritize candidate compounds for synthesis and characterization~\cite{schmidt2019,trezza2023leveraging,mancardi2023computational,mishra2023impact}.
Large-scale initiatives, such as the graph-network-based GNoME project~\cite{merchant_scaling_2023}, have largely expanded the number of predicted stable inorganic crystals, shifting the bottleneck from generating candidates to selecting those that are relevant for specific energy applications and for which predictions can be trusted.
In particular, the Energy-GNoME project~\cite{DEANGELIS2025100605} introduced a living, AI-driven screening protocol designed to extract energy-relevant candidates from Google DeepMind's GNoME database, which contains more than 380,000 newly predicted stable crystal structures. The resulting Energy-GNoME database includes more than 38,500 candidates for thermoelectric, photovoltaic, and battery applications. Rather than applying property-prediction models indiscriminately to the entire GNoME chemical space~\cite{merchant_scaling_2023}, the protocol first identifies those materials lying within a domain where predictions learned from specialized datasets can reasonably be expected to remain reliable. In the thermoelectric case, this procedure identifies unique candidate compositions for which the thermoelectric figure of merit \(zT\) is subsequently predicted~\cite{DEANGELIS2025100605}. 

A key methodological ingredient of Energy-GNoME is the concept of cross-domain data bias introduced in Ref.~\cite{Trezza2025}. Specialized materials databases are not unbiased samples of the full materials space, since the compounds investigated experimentally or computationally are typically selected according to prior scientific knowledge, chemical intuition, synthesis feasibility, and expected functionality. Consequently, a model that performs well when trained and tested within such a specialized dataset may still provide unreliable predictions when applied to materials originating from a substantially different domain. Reference~\cite{Trezza2025} addresses this issue by training binary classifiers in which materials from the specialized dataset are assigned to class 1, whereas samples drawn from a broader, less specialized database such as Materials Project~\cite{Jain2013MaterialsProject} are assigned to class 0. A skilled classifier therefore identifies the localized region of materials space represented by the specialized dataset and can be used to filter out materials for which subsequent regression would require unsafe extrapolation. In~\cite{DEANGELIS2025100605} this idea is operationalized through a committee of binary ML classifiers, termed AI-experts, which estimate whether previously unseen GNoME materials share the characteristic feature-space biases of the application-specific training set. Only materials passing this classification-based screening are forwarded to the regression stage for property prediction. This classification step is particularly relevant to the present work, since it defines a scientifically meaningful binary-learning problem: determining whether an inorganic material belongs to a localized functional-material domain or to the broader materials space. Here, we investigate whether quantum classification algorithms can perform this discrimination, thereby reformulating as a quantum-learning task a binary classification problem that already constitutes a central component of the Energy-GNoME materials-discovery workflow.

Quantum Machine Learning (QML) is emerging as a promising framework for identifying complex patterns in data~\cite{Biamonte2017}, with potentially relevant applications to the screening and discovery of energy materials. Moving beyond classical algorithms in the initial screening stage towards quantum-enhanced frameworks requires a paradigm shift in the representation and processing of complex materials descriptors. A key question is therefore whether the additional effort associated with this transition is justified by improved classification capabilities. In this regard, two properties appear particularly promising: (i) the high dimensionality of quantum-state Hilbert spaces, which may facilitate class separation by embedding the data into a richer feature space; and (ii) the nonlinear feature transformations induced by data encoding, quantum-circuit evolution, and measurement projections, which may capture complex higher-order dependencies among the input features~\cite{nielsen_quantum_2010}. Although quantum-state evolution is linear in Hilbert space, the resulting expectation values can depend nonlinearly on the classical input features through the parameterized encoding maps, the composition of quantum gates, and the measurement process.

Two quantum resources are particularly relevant to the expressive capabilities of such transformations: entanglement and non-stabilizerness, the latter commonly referred to as \textit{magic}. Entanglement enables nonseparable correlations among different qubits and, consequently, among the encoded features. Magic, on the other hand, quantifies the departure from the class of stabilizer states generated by Clifford circuits, which can otherwise be efficiently simulated classically. Non-Clifford operations capable of generating magic may therefore enrich the family of quantum states and feature transformations accessible to the model~\cite{leone_stabilizer_2022}. Nevertheless, neither entanglement nor magic alone guarantees improved learning or classification performance: their practical usefulness also depends on the encoding strategy, circuit architecture, trainability, measurement scheme, and structure of the input data. Entanglement and magic may be jointly used to characterize families of quantum states in a two-dimensional resource plane, thereby providing a compact representation of their quantum-resource content and expressibility~\cite{Maronese2026}.

QML methods are attracting increasing interest in physics, chemistry, and materials science, where they may support the identification of intricate structure--property relationships and accelerate materials screening and discovery~\cite{Sajjan2022}. Despite these promising characteristics, the practical advantage of QML over classical approaches remains an open question~\cite{cerezo_challenges_2022}. Available evidence indicates that well-established, off-the-shelf classical machine-learning models can outperform quantum classifiers; consequently, any potential quantum advantage must be rigorously assessed for the specific application under consideration~\cite{bowles2024betterclassicalsubtleart}. In this work, we investigate the classification of thermoelectric materials, a task previously addressed using classical classifiers within the Energy-GNoME project~\cite{DEANGELIS2025100605}.

Regarding the high dimensionality of quantum-state Hilbert spaces as a potential resource for data discrimination, our approach was inspired by the recent work in Ref.~\cite{DAmore2026}, which compares Projected Quantum Kernels (PQKs) with conventional Quantum Kernels (QKs) and classical Support Vector Machines (SVMs). In the PQK framework, a set of observables is measured on the encoded quantum state, and the resulting expectation values are subsequently used to construct the kernel. Here, we adopt a simplified and more direct strategy: rather than using these measurement outcomes to evaluate a kernel, we employ them directly as a new set of projected features for a downstream classical classifier. We refer to this approach as the Projected Quantum Features (PQF) method, which essentially implements a quantum-enhanced feature-transformation step.

Regarding the nonlinear transformations induced by quantum circuits, one of the most prominent examples is provided by Variational Quantum Classifiers (VQCs)~\cite{cerezo_variational_2021}. The form and expressive power of this nonlinearity depend on the circuit architecture, including the data-encoding strategy, the parametrized gates, and the entanglement topology~\cite{Brozzi2026}. In general, the resulting decision function may be interpreted as a complex trigonometric polynomial of the encoded features. Although this quantum-mechanical form of nonlinearity may be beneficial for specific classes of problems \cite{Sestini2026, 10606260}, there is no compelling reason to expect it to be universally advantageous for the classification of arbitrary classical data. The second idea explored in this work is therefore to adopt classical logistic regression as a baseline classifier and use it as an anchor upon which a trainable VQC correction is constructed. In this framework, the quantum-induced nonlinear contribution is tasked only with learning the deviations of the data from the predictions of the logistic-regression model, rather than learning the entire decision function from scratch. We refer to this hybrid approach as the logistic-regression-anchored variational quantum classifier (LRA-VQC), hereafter abbreviated as LRA whenever no ambiguity arises.

In summary, this work investigates two potential resources offered by quantum models---the high dimensionality of quantum-state Hilbert spaces and the nonlinear feature transformations induced by quantum circuits---and assesses their practical relevance to the classification of energy materials through a representative case study on thermoelectric compounds. To this end, we evaluate conventional VQC architectures and introduce two original approaches: the Projected Quantum Features (PQF) method, which exploits quantum measurements as transformed features, and the logistic-regression-anchored variational quantum classifier (LRA-VQC), which uses a VQC to model nonlinear deviations from a classical logistic-regression baseline. These three approaches are complementary and can be derived within a unified overarching theoretical framework, in which their differences arise primarily from whether the trainable parameters are assigned to the quantum state, the observable, or both.

The remainder of this article is organized as follows. Section~\ref{sec:problem} introduces the problem of classifying thermoelectric materials. Section~\ref{sec:overarching} presents the overarching theoretical framework, with particular emphasis on how trainable parameters are assigned to the quantum state, the observable, or both. Section~\ref{sec:encoding} describes the quantum data-encoding strategies shared by the different approaches. Section~\ref{sec:projected} introduces the Projected Quantum Features (PQF) model, whereas Section~\ref{sec:VQC} summarizes the essential components of the conventional VQC approach. Section~\ref{sec:LRA-VQC} then presents the logistic-regression-anchored variational quantum classifier (LRA-VQC). Section~\ref{sec:numerical-comparison} compares the predictive performance of the investigated models, including results obtained on real quantum hardware. Finally, Section~\ref{sec:conclusions} summarizes the main conclusions, while Appendix~\ref{sec:rotations} provides additional details on single-qubit rotations and the Bloch-sphere representation.

\section{Thermoelectric materials screening}
\label{sec:problem}

\subsection{Energy-GNoME}
\label{sec:egnome}

Following the procedure described in Ref.~\cite{DEANGELIS2025100605}, the starting dataset was constructed by combining a positive class of experimentally known thermoelectric materials with a negative class extracted from the Materials Project. After normalization of the chemical formulae and consolidation of repeated measurements, the positive class comprises 869 unique thermoelectric materials and 5,061 material--temperature observations. Indeed, a single material may occur in several rows because its thermoelectric properties were measured at different temperatures. The negative class was obtained from Materials Project compositions that do not overlap with the positive class and contains 944 unique materials represented by 5,736 temperature-dependent observations. The resulting full dataset, denoted by $\mathcal{D}_{\mathrm{full}}$, therefore contains $N_{\mathrm{full}}=5{,}061+5{,}736=10{,}797$ samples associated with 1,813 unique chemical compositions. Each sample corresponds to a specific material--temperature pair rather than necessarily to a distinct material. In the version adopted in this work, each sample is represented by $N_{\mathrm{F}}=145$ input features, including temperature.

Starting from $\mathcal{D}_{\mathrm{full}}$, we intentionally restricted the development dataset to $N_{\mathrm{tr+val}}=1{,}000$ samples in order to investigate the models in a data-limited training regime. These samples were divided into a training set with $N_{\mathrm{train}}=700$ samples and a validation set with $N_{\mathrm{valid}}=300$ samples. Both subsets were class-balanced and mutually exclusive at the material level: all temperature-dependent observations retained for a given chemical composition were assigned to only one of the two subsets. More specifically, the training and validation sets contain 557 and 230 unique materials, respectively.

To obtain a genuinely held-out material-wise test set, every row associated with a chemical composition appearing in either the training or validation set was excluded from the test set, including temperature-dependent rows that were not among the initial 1,000 development samples. The final test set therefore contains $N_{\mathrm{test}}=5{,}799$ samples associated with 1,026 previously unseen materials. As a consequence, 3,998 additional rows of $\mathcal{D}_{\mathrm{full}}$ were intentionally left unused, because they correspond to further temperatures of materials already represented in the training or validation set. Hence, the dataset effectively used for model development and evaluation is
\begin{equation}
\label{Dused}
\mathcal{D}_{\mathrm{used}}
=
\mathcal{D}_{\mathrm{train}}
\cup
\mathcal{D}_{\mathrm{valid}}
\cup
\mathcal{D}_{\mathrm{test}},
\end{equation}
with $N_{\mathrm{used}}=700+300+5{,}799=6{,}799$.
The remaining samples satisfy $N_{\mathrm{excluded}}=10{,}797-6{,}799=3{,}998$. This procedure prevents material leakage: no chemical composition represented during training or validation occurs in the final test set, even at a different temperature.

It is worth highlighting two characteristics of these datasets that are particularly relevant to their use in quantum-computing models. First, each raw feature is represented by a non-negative real number,
\begin{equation}
x_{\mathrm{R},\nu}\geq 0,
\qquad
\nu\in\{0,\ldots,N_{\mathrm{F}}-1\},
\end{equation}
where $\nu$ labels the individual features. This property facilitates some of the quantum data-encoding strategies introduced in Section~\ref{sec:encoding}. 
Second, the original number of features, $N_{\mathrm{F}}=145$, is too large for a practical implementation on currently available quantum hardware. 
We therefore employed SHAP (SHapley Additive exPlanations) \cite{Lundberg2017} to identify the $N_{\mathrm{F}}'=10<N_{\mathrm{F}}$ most informative features. 
To prevent information leakage, the SHAP ranking was determined using only the training data and was subsequently applied, without refitting, to the validation and test sets.

Before discussing the proposed quantum classifiers in detail, the following section briefly reviews the classical binary classifiers adopted as reference models.

\subsection{Logistic regression and other classifiers}
\label{sec:logistic}

We consider a supervised binary-classification problem in which each sample is represented by a feature vector $\mathbf{x}_{\mathrm{R}}\in\mathbb{R}^{N_{\mathrm{F}}}$ and associated with a class label $y\in\{0,1\}$. Given a training set
\begin{equation}
\label{eqn:training-set}
\mathcal{D}_{\mathrm{train}}
=
\left\{
\left(\mathbf{x}^{(i)}_{\mathrm{R}},y^{(i)}\right)
\right\}_{i=0}^{N_{\mathrm{train}}-1},
\end{equation}
the objective is to learn a parameterized mapping
\begin{equation}
\label{eqn:mapping}
f_{\bm{\theta}}:
\mathbb{R}^{N_{\mathrm{F}}}\rightarrow[0,1],
\qquad
\mathbf{x}_{\mathrm{R}}\mapsto
p_{\bm{\theta}}(y=1\mid\mathbf{x}_{\mathrm{R}}),
\end{equation}
where $p_{\bm{\theta}}(y=1\mid\mathbf{x}_{\mathrm{R}})$ is the estimated probability that $\mathbf{x}_{\mathrm{R}}$ represents a thermoelectric material. The subscript in $\mathbf{x}_{\mathrm{R}}$ means that the features are raw, namely with original physical intervals. Before training, the input features are typically rescaled to prevent differences in their physical units and numerical ranges from disproportionately affecting the classification model. A common choice is standardization, which transforms each feature using the mean and standard deviation of the training set:
\begin{equation}
\label{eqn:standard-scaling}
\widetilde{x}_\nu^{\,\mathrm{std}}=
\frac{x_{\mathrm{R},\nu}-\mu_\nu^{\mathrm{train}}}
{\sigma_\nu^{\mathrm{train}}},
\end{equation}
where $\nu\in[0,\,N_{\mathrm{F}}-1]$ identifies the generic feature. In quantum computing, other options are preferable (see Section~\ref{sec:encoding} for details). In the following, $\mathbf{x}_{\mathrm{R}}$ denotes the raw feature vector, whereas $\mathbf{x}$ denotes the corresponding preprocessed vector obtained after applying the normalization or standardization required by the model. For notational simplicity, the specific preprocessing transformation is left implicit whenever it is clear from the context.

Putting aside the specific rescaling performed during the normalization step, the predicted label is obtained by applying a decision threshold $\tau$:
\begin{equation}
\label{eqn:threshold}
y_{\bm{\theta}}(\mathbf{x})
=
\begin{cases}
1, & p_{\bm{\theta}}(y=1\mid\mathbf{x})\geq\tau,\\
0, & p_{\bm{\theta}}(y=1\mid\mathbf{x})<\tau.
\end{cases}
\end{equation}
Unless otherwise specified, the conventional threshold $\tau=0.5$ is adopted. The model parameters $\bm{\theta}$ are determined by minimizing a suitable empirical loss over the training set, while predictive performance is assessed on previously unseen samples.

For binary \emph{logistic regression}, the probability of the positive class is modeled as
\begin{equation}
\label{eqn:binary}
p_{\mathrm{L}}(y=1\mid\mathbf{x})
=
\eta_{\,\mathrm{L}}\!\left(
s_{\mathrm{L}}(\mathbf{x})
\right),
\end{equation}
where
\begin{equation}
\label{eqn:logit}
s_{\mathrm{L}}(\mathbf{x})
=
b+\mathbf{w}^{\mathsf{T}}\mathbf{x}
=
b+\sum_{\nu=0}^{N_{\mathrm{F}}-1}w_\nu x_\nu
\end{equation}
is the classification score, or logit, and
\begin{equation}
\label{eqn:sigmoid}
\eta_{\,\mathrm{L}}(s)
=
\frac{1}{1+\exp(-s)}
\end{equation}
is the sigmoid function. The model parameters are collected in $\bm{\theta}=(b,\mathbf{w})\in\mathbb{R}^{N_{\bm{\theta}}^{\mathrm{(L)}}}$, where $N_{\bm{\theta}}^\mathrm{(L)}=N_{\mathrm{F}}+1$, while $b\in\mathbb{R}$ is the bias and $\mathbf{w}\in\mathbb{R}^{N_{\mathrm{F}}}$ is the vector of feature weights. The complete mapping can therefore be written as
\begin{equation}
\label{eqn:mapping-binary}
f_{\mathrm{L}}:
\mathbb{R}^{N_{\mathrm{F}}}
\to
[0,1],
\qquad
\mathbf{x}
\mapsto
\eta_{\,\mathrm{L}}\!\left(
b+\mathbf{w}^{\mathsf{T}}\mathbf{x}
\right).
\end{equation}
Using the conventional threshold $\tau=0.5$, taking into account that $\eta_{\,\mathrm{L}}(s)\geq 0.5$ if and only if $s\geq 0$, the decision rule can be expressed as
\begin{equation}
\label{eqn:decision-rule}
y_{\,\mathrm{L}}(\mathbf{x})
=
\begin{cases}
1, & b+\mathbf{w}^{\mathsf{T}}\mathbf{x}\geq0,\\
0, & b+\mathbf{w}^{\mathsf{T}}\mathbf{x}<0.
\end{cases}
\end{equation}
Consequently, logistic regression defines the linear decision boundary
\begin{equation}
\label{eqn:decision-boundary}
b+\mathbf{w}^{\mathsf{T}}\mathbf{x}=0
\end{equation}
in the original feature space. The parameters are determined by minimizing the $L_2$-regularized binary cross-entropy loss with possibly a penalty for ensuring some regularization, namely
\begin{eqnarray}
\label{eqn:loss-binary}
\mathcal{L}(\bm{\theta})
=&&
-\frac{1}{N_{\mathrm{train}}}
\sum_{i=0}^{N_{\mathrm{train}}-1}
\left[
y_i\log p_i
+
(1-y_i)\log(1-p_i)
\right]\nonumber\\
&&
+\sum_{j=0}^{N_{\bm{\theta}}-1}\lambda_j\theta_j^2,
\end{eqnarray}
where $y_i\in\{0,1\}$ is the ground-truth binary label
associated with the $i$-th training sample,
$p_i=p_{\mathrm{L}}(y=1\mid\mathbf{x}_i)$ is the
corresponding predicted probability of class $1$, $N_{\bm{\theta}}=N_{\bm{\theta}}^\mathrm{(L)}=N_{\mathrm{F}}+1$ and
$\lambda_j\geq 0$ control the strength of the regularization. In case of logistic regression, typically $\lambda_0=0$ and $\lambda_j=\lambda_{\mathrm{w}}$ for all $1\leq j\leq N_{\bm{\theta}}-1$. The regularization term discourages excessively large coefficients and can improve generalization to previously unseen samples.

More flexible classical models can be considered when the linear decision boundary of logistic regression is insufficient. One possibility is \emph{polynomial logistic regression}, in which the original feature vector $\mathbf{x}$ is replaced by an expanded representation containing monomials and feature-interaction terms up to a certain degree. Although this approach preserves the probabilistic interpretation of logistic regression, the number of expanded features grows rapidly with both $N_{\mathrm{F}}$ and the polynomial degree, increasing the computational cost and the risk of overfitting. A more powerful alternative for tabular data is \emph{eXtreme Gradient Boosting} (XGBoost), which constructs an additive ensemble of decision trees by sequentially correcting the errors of the preceding trees while minimizing a regularized objective~\cite{Chen2016}. XGBoost can capture nonlinear thresholds and high-order feature interactions without requiring an explicit polynomial expansion and is widely regarded as a strong state-of-the-art baseline for heterogeneous tabular datasets. Polynomial logistic regression and XGBoost therefore provide increasingly flexible classical references against which the benefits of the quantum classification models can be assessed.

\section{Overarching theoretical framework}
\label{sec:overarching}

The purpose of this section is to outline the key concepts underlying the proposed quantum models, while leaving the technical details to the subsequent sections. Let us suppose we load, or encode, a classical feature vector $\mathbf{x}$ into a quantum state $\ket{\psi(\mathbf{x})}$ by using $n$ qubits (see Section~\ref{sec:encoding} for details). Next, the key point for constructing a quantum classifier is the definition of a quantum score $s_{\mathrm{Q}}$, given by the expectation value of a suitable observable $O$:
\begin{equation}
\label{eqn:quantum-score}
s_{\mathrm{Q}}(\mathbf{x})
=
\braket{\psi(\mathbf{x})|O|\psi(\mathbf{x})}.
\end{equation}
The corresponding probability of assigning the input $\mathbf{x}$ to class $y=1$ is then defined as
\begin{equation}
\label{eqn:binary-quantum}
p_{\mathrm{Q}}(y=1\mid\mathbf{x})
=
\eta_{\,\mathrm{Q}}\!\left(
s_{\mathrm{Q}}(\mathbf{x})
\right),
\end{equation}
where the choice of the mapping $\eta_{\,\mathrm{Q}}$ depends on the range of the quantum score. 

Across the different classes of quantum models, a central design choice in constructing a variational architecture concerns how the learnable parameters $\bm{\theta}$ are allocated. They may enter the observable $O$, the quantum state $\ket{\psi(\mathbf{x})}$, or both. These alternative parameterizations and their implications will be discussed in the following sections.

\subsection{Projected quantum feature (PQF) model}

In this design, the learnable parameters $\bm{\theta}$ are allocated to the observable only, namely
\begin{equation}
\label{eqn:PQF-observable}
O_K(\bm{\theta})=\sum_{p=0}^{N_K}
\theta_p\,\sigma_p=
b_K+\sum_{p=1}^{N_K}
w_{K,\,p}\,\sigma_p,
\end{equation}
where $\sigma_p$ is an operator which can be
expressed as a tensor product of single-qubit Pauli operators (see Section~\ref{sec:projected} for details). Defining a quantum score as the expectation value of the previous observable yields
\begin{equation}
\label{eqn:PQF-quantum-score}
s_{K}(\mathbf{x})
=\braket{\psi_{\mathrm{E}}(\mathbf{x})|O_{K}(\bm{\theta})|\psi_{\mathrm{E}}(\mathbf{x})}
=b_K+
\mathbf{w}_{K}^{\mathsf{T}}
\mathbf{z}_{K}(\mathbf{x})
\end{equation}
where $\ket{\psi_{\mathrm{E}}(\mathbf{x})} = U_{\mathrm{E}}(\mathbf{x})\ket{0}^{\otimes n}$ is the quantum state which encodes the input data features (see Section~\ref{sec:encoding} for details) and $\mathbf{z}_{K}(\mathbf{x})$ is a vector of \textit{projected quantum features} (PQFs). Essentially, Eq. (\ref{eqn:PQF-quantum-score}) defines a quantum logit for this model, which can be used to feed the sigmoid function given by Eq. (\ref{eqn:sigmoid}), i.e. $p_{K}(y=1\mid\mathbf{x})=\eta_{\,\mathrm{L}}\!\left(s_{K}(\mathbf{x})\right)$. Hence PQFs are directly treated as quantum-transformed features and subsequently processed by a classical classifier. In case $N_K>N_{\mathrm{F}}$, the underlying expectation is that embedding the original data into a higher-dimensional feature space may render the classes more easily separable and, consequently, improve classification performance. We refer to this approach as the projected quantum feature (PQF) model (see Section~\ref{sec:projected}).

\subsection{Variational quantum classifier (VQC)}

In this design, the learnable parameters $\bm{\theta}$ are allocated in the quantum state only, namely
\begin{equation}
\label{eqn:VQC-state}
\ket{\psi(\mathbf{x},\bm{\theta})}=
V_{\mathrm{A}}(\bm{\theta})\,
\ket{\psi_{\mathrm{E}}(\mathbf{x})}=
V_{\mathrm{A}}(\bm{\theta})\,
U_{\mathrm{E}}(\mathbf{x})\,\ket{0}^{\otimes n},
\end{equation}
where $V_{\mathrm{A}}(\bm{\theta})$ is the gate with the learnable parameters $\bm{\theta}$ (see Section~\ref{sec:VQC} for details). A quantum score can be defined in this model as
\begin{equation}
\label{eqn:QVC-quantum-score}
s_{\mathrm{V}}(\mathbf{x})=
\braket{\psi(\mathbf{x},\bm{\theta})|O|\psi(\mathbf{x},\bm{\theta})},
\end{equation}
where $O=Z^{\otimes n}$. An alternative formulation consists in using the observable $O_0'=I^{\otimes (n-1)}\otimes Z\equiv Z_0$ and $Z_0$ denotes the Pauli-$Z$ operator acting on qubit $0$, which is the least significant qubit (LSQb). Essentially, Eq. (\ref{eqn:QVC-quantum-score}) defines a quantum score $s_{\mathrm{V}}\in[-1,\,+1]$ for this model, which can be used as
\begin{equation}
\label{eqn:QVC-binary}
p_{\mathrm{V}}(y=1\mid\mathbf{x})
=
\eta_{\,\mathrm{V}}\!\left(
s_{\mathrm{V}}(\mathbf{x})
\right),\qquad
\eta_{\,\mathrm{V}}(s)=\frac{1-s}{2}.
\end{equation}
In Section~\ref{sec:VQC}, different implementations of this idea are discussed, mainly depending on the adopted gate $V_{\mathrm{A}}(\bm{\theta})$.

\subsection{Logistic-regression-anchored (LRA) VQC}

In this design, some learnable parameters $\bm{\theta}_{\mathrm{L}}$, computed by a classical model, are added as an anchor. The learnable parameters $\bm{\theta}_{\mathrm{L}}$ and $\bm{\theta}$ are allocated to the observable and the quantum state respectively, namely
\begin{equation}
\label{eqn:quantum-logit-LRA}
s_{\mathrm{A}}(\mathbf{x})=
\braket{\psi(\mathbf{x},\bm{\theta})|O_{\mathcal{E}}(\bm{\theta}_{\mathrm{L}})|\psi(\mathbf{x},\bm{\theta})},
\end{equation}
where the same state construction introduced for the VQC applies. In
particular, the trainable quantum state is defined as
$\ket{\psi(\mathbf{x},\bm{\theta})}
=V_{\mathcal{A}}(\bm{\theta})
\ket{\psi_{\mathcal{E}}(\mathbf{x})}$, while the encoded state is
given by $\ket{\psi_{\mathcal{E}}(\mathbf{x})}=
U_{\mathcal{E}}(\mathbf{x})
\ket{0}^{\otimes n}$. The feature mapping $U_{\mathcal{E}}(\mathbf{x})$, among all possible $U_{\mathrm{E}}(\mathbf{x})$, must ensure some simple extraction of the features $\mathbf{x}$ (see Section~\ref{sec:LRA-VQC} for details). The observable $O_{\mathcal{E}}(\bm{\theta}_{\mathrm{L}})$ must be designed complementary to the encoding such that, in case of pure encoding only, 
\begin{equation}
\label{eqn:quantum-logit-LRA-observable}
\braket{\psi_{\mathcal{E}}(\mathbf{x})
|O_{\mathcal{E}}(\bm{\theta}_{\mathrm{L}})|
\psi_{\mathcal{E}}(\mathbf{x})}
=b_{\mathrm{L}}+
\mathbf{w}_{\mathrm{L}}^{\mathsf{T}}\mathbf{x},
\end{equation}
which is essentially the linear decision boundary given by the logistic regression model, reported in Eq. (\ref{eqn:decision-boundary}). Once $U_{\mathcal{E}}(\mathbf{x})$ and $O_{\mathcal{E}}(\bm{\theta}_{\mathrm{L}})$ are defined, then the variational ansatz $V_{\mathcal{A}}(\bm{\theta})$ can be added on top and, among all possible $V_{\mathrm{A}}(\bm{\theta})$, it must be designed under the special condition that $V_{\mathcal{A}}(\bm{0})=I^{\otimes n}$, which ensures the proper \textit{anchoring} to the previous logistic regression model. The output is given by $p_{\mathrm{A}}(y=1\mid\mathbf{x})=\eta_{\,\mathrm{L}}\!\left(s_{\mathrm{A}}(\mathbf{x})\right)$, where $\eta_{\,\mathrm{L}(s)}$ is the same sigmoid function given by Eq. (\ref{eqn:sigmoid}).

As a concluding remark, for simplicity, the preceding derivation of the anchored model has been presented with reference to logistic regression, whose decision boundary is expressed in terms of the original input features $\mathbf{x}$. The same construction can be readily extended to anchor the model to a decision boundary defined in terms of the projected quantum features $\mathbf{z}_K(\mathbf{x})$ introduced in Eq.~(\ref{eqn:PQF-quantum-score}).

\section{Quantum data encoding}
\label{sec:encoding}

The construction of a quantum classifier begins by loading, or encoding, a classical feature vector $\mathbf{x}$ into a quantum state $\ket{\psi_{\mathrm{E}}(\mathbf{x})}$ belonging to a Hilbert space of dimension $N=2^n$, where $n$ denotes the number of qubits. The number of qubits $n$ required to represent a classical feature vector depends on the adopted encoding strategy. The exponentially growing dimensionality of this space is an intrinsic feature of quantum mechanics and may also provide computational advantages by enabling rich representations of classical data~\cite{nielsen_quantum_2010}. 

Throughout this paper, the quantum state produced by the encoding circuit is denoted by $\ket{\psi_{\mathrm{E}}(\mathbf{x})}=U_{\mathrm{E}}(\mathbf{x})\ket{0}^{\otimes n}$, where the data-dependent unitary operator $U_{\mathrm{E}}(\mathbf{x})$ may correspond to any of the encoding circuits introduced in the following sections, i.e. 
\begin{equation}
\label{eqn:type-encoding}
U_{\mathrm{E}}(\mathbf{x})\in\{U_{\varphi},\,U_{\varphi'},U_{\varphi\phi},\,U_{\mathrm{A}},\,U_{\mathrm{A}'},\,U_{\mathrm{A}''}\}.
\end{equation}
The encoding schemes considered in this work are described below. For the sake of simplicity, the concept of re-uploading~\cite{PerezSalinas2020datareuploading} is not discussed in the work, because it was not beneficial for the present dataset.

\subsection{Angle and dense-angle encodings}
\label{sec:angle}

Before introducing angle encoding, it is useful to recall that quantum data encoding may require preprocessing strategies different from those commonly adopted for the classical models discussed in Section~\ref{sec:logistic}. A frequently used choice is min--max scaling, which maps the values of each feature in the training set onto a prescribed interval:
\begin{equation}
\label{eqn:minmax-scaling}
\widetilde{x}_\nu
=
a+
(b-a)
\frac{x_{\mathrm{R},\nu}-x_{\nu,\min}^{\mathrm{train}}}
{x_{\nu,\max}^{\mathrm{train}}
-x_{\nu,\min}^{\mathrm{train}}},
\end{equation}
where $[a,\,b]$ denotes the selected target interval. Min--max scaling is particularly convenient for quantum angle encoding because the scaled values can be mapped directly onto a prescribed range of rotation angles. For example, one may define
\begin{equation}
\label{eqn:angle-scaling}
\varphi_\nu
=
\pi\widetilde{x}_\nu,
\qquad
\widetilde{x}_\nu
\in [a,\,b]\subseteq[0,\,1],
\end{equation}
or, alternatively,
\begin{equation}
\label{eqn:angle-scaling2}
\varphi_\nu'
=
\arccos\!\left(
\widetilde{x}'_\nu
\right),
\qquad
\widetilde{x}'_\nu
\in[a',\,b']=[-1,\,1].
\end{equation}

To prevent information leakage, all scaling parameters must be determined exclusively from the training set and subsequently applied, without refitting, to the validation and test sets. Thus, the scaling parameters inferred from the training data are retained unchanged when processing previously unseen samples. Consequently, if a validation or test value lies outside the range observed during training, its scaled value may fall outside the prescribed interval $[a,\,b]$. In the case of Eq.~(\ref{eqn:angle-scaling}), this may produce a limited number of negative rotation angles or angles exceeding the intended encoding range. Although such rotations remain mathematically and physically well defined, their periodicity causes them to represent an extrapolation of the chosen encoding and may introduce an unintended wrap-around effect. For this reason, the scaled values of previously unseen samples may be clipped to the interval $[a,\,b]$, thereby limiting wrap-around while preserving the encoding range established during training. Nevertheless, min--max scaling is not intrinsically required by quantum computing and may not be optimal for physical features that follow an approximately Gaussian distribution, for which standardization based on the mean and standard deviation may provide a more representative preprocessing strategy. In the following, for notational simplicity, $\mathbf{x}\in[0,\,1]^{N_{\mathrm{F}}}$ denotes the preprocessed feature vector used in the quantum models. When the features are instead scaled to the interval $[-1,\,1]$, the notation $\mathbf{x}'\in[-1,\,1]^{N_{\mathrm{F}}}$ is used to make this distinction explicit.

In conventional \textit{angle encoding}, the number of qubits required to encode $N_{\mathrm{F}}$ features is $n=N_{\mathrm{F}}$. Therefore, the feature index $\nu$ used above can equivalently be replaced by the qubit index $q$, since each feature is associated with a distinct qubit. More specifically, each feature may be encoded through a single-qubit rotation (see Appendix~\ref{sec:rotations}) according to
\begin{equation}
\label{eqn:angle-encoding}
U_{\varphi}(\mathbf{x})
=
\bigotimes_{q=0}^{n-1}
R_y^{(q)}
\!\left(
\pi\,x_q
\right),
\end{equation}
or, using the alternative transformation introduced in Eq.~(\ref{eqn:angle-scaling2}), which can be called \textit{cos-angle encoding}, as
\begin{equation}
\label{eqn:angle-encoding2}
U_{\varphi'}(\mathbf{x})
=
\bigotimes_{q=0}^{n-1}
R_y^{(q)}
\!\left[
\arccos\!\left(
x'_q
\right)
\right].
\end{equation}
Using the previous gates, it is possible to define $\ket{\psi_{\varphi}(\mathbf{x})}=U_{\varphi}(\mathbf{x})\ket{0}^{\otimes n}$ and $\ket{\psi_{\varphi'}(\mathbf{x})}=U_{\varphi'}(\mathbf{x})\ket{0}^{\otimes n}$, respectively. 

In \textit{dense-angle encoding}, two features are instead encoded on each qubit, reducing the number of required qubits to
$n=\left\lceil N_{\mathrm{F}}/2\right\rceil$. If $N_{\mathrm{F}}$ is odd, the missing component associated with the final qubit must be assigned a fixed padding value. The index $q$ therefore identifies a pair of features, which may be encoded through two consecutive rotations (see Appendix~\ref{sec:rotations}) about different axes:
\begin{equation}
\label{eqn:dense-angle-encoding}
U_{\varphi\phi}(\mathbf{x})
=
\bigotimes_{q=0}^{n-1}
R_z^{(q)}
\!\left(
2\pi\,x_{2q+1}
\right)
R_y^{(q)}
\!\left(
\pi\,x_{2q}
\right).
\end{equation}
Using the previous gates, it is possible to define $\ket{\psi_{\varphi\phi}(\mathbf{x})}=U_{\varphi\phi}(\mathbf{x})\ket{0}^{\otimes n}$.

\subsection{Amplitude and probability encodings}
\label{sec:amplitude}

Three closely related encoding strategies can then be distinguished here. In conventional \textit{amplitude encoding}, the normalized feature components are directly assigned to the probability amplitudes of the computational basis states:
\begin{equation}
\label{eqn:amplitude-encoding}
\ket{\psi_{\mathrm{A}}(\mathbf{x})}=
U_{\mathrm{A}}(\mathbf{x})\ket{0}^{\otimes n}
=
\sum_{\nu=0}^{N_{\mathrm{F}}-1}
\widehat{x}_{\nu}\ket{\nu},
\end{equation}
where $\widehat{x}_{\nu}=x_\nu/M(\mathbf{x})$ and
\begin{equation}
\label{eqn:magnitude}
M(\mathbf{x})
=\lVert\mathbf{x}\rVert_2=
\sqrt{\sum_{\nu=0}^{N_{\mathrm{F}}-1}
x^2_\nu}.
\end{equation}
Using the previous quantum state, it is possible to define a gate $U_{\mathrm{A}}(\mathbf{x})$ such that $\ket{\psi_{\mathrm{A}}(\mathbf{x})}=U_{\mathrm{A}}(\mathbf{x})\ket{0}^{\otimes n}$. The number of required qubits is therefore
\begin{equation}
\label{eqn:amplitude-qubits}
n
=
\left\lceil
\log_2 N_{\mathrm{F}}
\right\rceil,
\end{equation}
and the amplitudes associated with the remaining
$2^n-N_{\mathrm{F}}$ computational basis states are set to zero. This encoding can represent both positive and negative real-valued features. However, because the state is invariant under multiplication of the original feature vector by an arbitrary positive scalar, its Euclidean magnitude is not retained.

A second possibility is \textit{probability encoding}, which requires nonnegative features $x_\nu\geq 0$, as may be obtained, for example, through min--max scaling. Defining the sample-dependent normalization factor
\begin{equation}
\label{eqn:probability-normalization}
S(\mathbf{x})
=\lVert\mathbf{x}\rVert_1=
\sum_{\nu=0}^{N_{\mathrm{F}}-1}
\left|x_\nu\right|=
\sum_{\nu=0}^{N_{\mathrm{F}}-1}
x_\nu,
\end{equation}
the normalized features are interpreted as probabilities, whereas their square roots provide the corresponding quantum amplitudes:
\begin{equation}
\label{eqn:probability-encoding}
\ket{\psi_{\mathrm{A}'}(\mathbf{x})}=
U_{\mathrm{A}'}(\mathbf{x})\ket{0}^{\otimes n}
=
\sum_{\nu=0}^{N_{\mathrm{F}}-1}
\sqrt{
\frac{x_\nu}
{S(\mathbf{x})}
}
\ket{\nu}.
\end{equation}
Accordingly, a measurement in the computational basis returns the probability
\begin{equation}
\label{eqn:probability-encoding-measurement}
\left|
\braket{\nu|\psi_{\mathrm{A}'}(\mathbf{x})}
\right|^2
=
\frac{x_\nu}
{S(\mathbf{x})}.
\end{equation}
This construction also requires
$n=\left\lceil\log_2N_{\mathrm{F}}\right\rceil$ qubits and zero padding when $N_{\mathrm{F}}$ is not a power of two. Unlike conventional amplitude encoding, the encoded amplitudes are nonnegative and their squared values are directly proportional to the original features. Nevertheless, the sample-dependent factor $S(\mathbf{x})$ is removed by normalization, so that feature vectors differing only by an overall positive factor are mapped onto the same quantum state.

The previous two strategies mix the features by means of the magnitude $M(\mathbf{x})$ defined by Eq. (\ref{eqn:magnitude}) or the size $S(\mathbf{x})$ given by Eq. (\ref{eqn:probability-normalization}) and this could make the classification more complex. In order to avoid this, a third strategy consists in \textit{probability encoding with size} by introducing an additional auxiliary basis state. Let $S_{\max}$ be a fixed constant satisfying
\begin{equation}
\label{eqn:auxiliary-bound}
S(\mathbf{x})
=
\sum_{\nu=0}^{N_{\mathrm{F}}-1}
x_\nu
\leq S_{\max}
\end{equation}
for every sample. The feature probabilities can then be defined without sample-wise normalization as
\begin{equation}
\label{eqn:auxiliary-feature-probabilities}
\frac{x_\nu}{S_{\max}},
\qquad
\nu=0,\ldots,N_{\mathrm{F}}-1,
\end{equation}
while the residual probability is assigned to an auxiliary state:
\begin{equation}
\label{eqn:auxiliary-probability}
1-
\frac{S(\mathbf{x})}{S_{\max}}.
\end{equation}
The resulting normalized quantum state is
\begin{eqnarray} 
\label{eqn:auxiliary-state-encoding}
\ket{\psi_{\mathrm{A}''}(\mathbf{x})}
&=&U_{\mathrm{A}''}(\mathbf{x})\ket{0}^{\otimes n}\nonumber\\
&=&
\sum_{\nu=0}^{N_{\mathrm{F}}-1}
\sqrt{
\frac{x_\nu}{S_{\max}}
}
\ket{\nu}
+
\sqrt{
1-\frac{S(\mathbf{x})}{S_{\max}}
}
\ket{N_{\mathrm{F}}}.
\end{eqnarray}
In this case, the quantity $S(\mathbf{x})$ is retained through the total probability assigned to the feature subspace, while the auxiliary-state probability stores the complementary fraction. The number of required qubits becomes
\begin{equation}
\label{eqn:auxiliary-qubits}
n
=
\left\lceil
\log_2\!\left(N_{\mathrm{F}}+1\right)
\right\rceil,
\end{equation}
with any remaining computational basis states again assigned zero amplitude. If every scaled feature belongs to $[a,\,b]\subseteq[0,\,1]$, a convenient sample-independent choice is $S_{\max}=N_{\mathrm{F}}$. More generally, $S_{\max}$ must be determined from the training data alone and retained unchanged for validation, test, and previously unseen samples. This auxiliary-state construction therefore preserves both the relative composition of the feature vector and its $L_1$ magnitude, at the cost of one additional basis state and, in some cases, one additional qubit.

\section{Projected quantum feature (PQF) model}
\label{sec:projected}

This design is called projected quantum feature (PQF) model because only a selected subset of the matrices of the generic expansion of the observable is retained, thereby reducing both measurement and computational costs. 

The projected quantum features are defined as expectation values of selected observables drawn from the set of Pauli strings. In general, the set of $n$-qubit Pauli strings
\begin{equation}
\label{eqn:pauli-space}
\mathcal{P}_n
=
\{I,X,Y,Z\}^{\otimes n}
\end{equation}
contains $N_{\mathrm{P}}=4^n=N^2$ operators and forms an orthogonal basis for the vector space of linear operators acting on the $N=2^n$ dimensional Hilbert space~\cite{nielsen_quantum_2010}. The single-qubit Pauli basis is defined by
\begin{equation}
\label{eqn:pauli-basis}
\begin{aligned}
I&=
\begin{pmatrix}
1&0\\
0&1
\end{pmatrix},
&
X&=
\begin{pmatrix}
0&1\\
1&0
\end{pmatrix},
\\[1mm]
Y&=
\begin{pmatrix}
0&-\mathrm{i}\\
\mathrm{i}&0
\end{pmatrix},
&
Z&=
\begin{pmatrix}
1&0\\
0&-1
\end{pmatrix}.
\end{aligned}
\end{equation}
Each operator $\sigma_p\in\mathcal{P}_n$ in this set has dimension $2^n\times2^n$ and can be expressed as a tensor product of single-qubit Pauli operators:
\begin{equation}
\label{eqn:Pauli-string}
\sigma_p
=
\bigotimes_{q=0}^{n-1}
\sigma_p^{(q)},
\qquad
\sigma_p^{(q)}
\in
\{I,X,Y,Z\}.
\end{equation}
Specifically, the Pauli strings satisfy
\begin{equation}
\label{eqn:Pauli-orthogonality}
\operatorname{Tr}(\sigma_p\sigma_q)
=
2^n\delta_{pq}.
\end{equation}
Because the set of the Pauli strings forms an orthogonal basis, it can be used to expand the operator $O$ in Eq. (\ref{eqn:quantum-score}) as
\begin{equation}
\label{eqn:Pauli-expansion}
O=
\sum_{p=0}^{N_{\mathrm{P}}-1}
\gamma_p\,\sigma_p,
\qquad
N_{\mathrm{P}}=4^n=N^2,
\end{equation}
where the expansion coefficients are given by
\begin{equation}
\label{eqn:Pauli-coefficients}
\gamma_p=
\frac{1}{2^n}
\operatorname{Tr}
\left(
\sigma_pO
\right).
\end{equation}
Because both $O$ and $\sigma_p$ are Hermitian, the coefficients $\gamma_p$ are real. Temporarily setting aside the learnable parameters $\bm{\theta}$, substituting Eq.~(\ref{eqn:Pauli-expansion}) into Eq.~(\ref{eqn:quantum-score}) yields
\begin{eqnarray} 
\label{eqn:fidelity-kernel3}
s_{\mathrm{Q}}(\mathbf{x})
&=&
\bra{\psi(\mathbf{x})}
\left[
\sum_{p=0}^{N_{\mathrm{P}}-1}
\gamma_p\,\sigma_p
\right]
\ket{\psi(\mathbf{x})}
\nonumber\\
&=&
\sum_{p=0}^{N_{\mathrm{P}}-1}
\gamma_p\,z_p(\mathbf{x}),
\end{eqnarray}
where
\begin{equation}
\label{eqn:Pauli-expectation}
z_p(\mathbf{x})
=
\bra{\psi(\mathbf{x})}
\sigma_p
\ket{\psi(\mathbf{x})}
\in\mathbb{R}
\end{equation}
is the expectation value of the Pauli observable $\sigma_p$. The complete representation $\mathbf{z}(\mathbf{x})\in\mathbb{R}^{N_{\mathrm{P}}}$ defined by Eq.~(\ref{eqn:Pauli-expectation}) can be interpreted as a quantum feature map
\begin{equation}
\label{eqn:mapping-features}
m_{\mathrm{P}}:
\mathbb{R}^{N_{\mathrm{F}}}
\to
\mathbb{R}^{N_{\mathrm{P}}},
\qquad
\mathbf{x}
\mapsto
\mathbf{z}(\mathbf{x}),
\end{equation}
where $N_{\mathrm{F}}$ is the number of original classical features and $N_{\mathrm{P}}=4^n$ is the number of $n$-qubit Pauli strings, including the identity.

Retaining the complete set of Pauli strings may be computationally prohibitive and experimentally impractical. Therefore, different truncation strategies can be introduced, for example by restricting the expansion to Pauli strings with a prescribed maximum weight. The Pauli weight of $\sigma_p$, denoted by $\operatorname{wt}(\sigma_p)$, is defined as the number of non-identity factors in the corresponding Pauli string. The Pauli strings can therefore be grouped according to their weight $k$, ranging from $k=0$ for the identity $I^{\otimes n}$ to $k=n$ for strings containing no identity factors. For a given weight $k$, the number of distinct Pauli strings is
\begin{equation}
\label{eqn:Nk}
n_k
=
3^k
\binom{n}{k}.
\end{equation}
Indeed, the binomial coefficient in the previous formula specifies the positions of the $k$ non-identity factors, while each selected position can independently contain one of the three operators $X$, $Y$, or $Z$. Consequently,
\begin{equation}
\sum_{k=0}^{n}n_k
=
\sum_{k=0}^{n}
3^k\binom{n}{k}
=
4^n
=
N_{\mathrm{P}}.
\end{equation}
These expectation values are not all statistically independent, because they originate from the same quantum state. The measurement cost nevertheless grows rapidly with $n$, motivating a truncated representation based on low-weight Pauli observables.

Let $K\leq n$ denote the maximum Pauli weight retained in the projected representation. Excluding the identity, whose expectation value is always equal to one and can therefore be absorbed into the intercept of the classical classifier, the number of nontrivial Pauli features up to weight $K$ is
\begin{equation}
\label{eqn:NPK}
N_{K}
=
\sum_{k=1}^{K}
n_k
=
\sum_{k=1}^{K}
3^k
\binom{n}{k}
\sim
\frac{3^K}{K!}\,n^K.
\end{equation}
For a fixed maximum Pauli weight $K$, the size of the truncated  observable set scales as a polynomial of degree $K$. In particular, for example: $N_1
=3n$, $N_2=3n\,(3n-1)/2$, $N_3=3n\,\left(3n^2-6n+5\right)/2$, etc. The corresponding truncated projected-feature vector is defined as
\begin{equation}
\label{eqn:truncated-feature-vector}
\mathbf{z}_{K}(\mathbf{x})
=
\left(
\bra{\psi(\mathbf{x})}
\sigma_p
\ket{\psi(\mathbf{x})}
\right)_{
\sigma_p\in\mathcal{P}_{K}^\circ
}
\in
\mathbb{R}^{N_{K}},
\end{equation}
where 
\begin{equation}
\label{eqn:truncated-Pauli-set}
\mathcal{P}_{K}^\circ
=
\left\{
\sigma\in\mathcal{P}_n
\,\middle|\,
1\leq\operatorname{wt}(\sigma)\leq K
\right\}
%=
%\left\{
%\sigma_p
%\right\}_{p=1}^{N_{K}}
\subset\mathcal{P}_n,
\end{equation}
which excludes the identity operator. Accordingly, the truncated quantum feature map is defined as
\begin{equation}
\label{eqn:mapping-features-truncated}
m_{K}:
\mathbb{R}^{N_{\mathrm{F}}}
\to
\mathbb{R}^{N_{K}},
\qquad
\mathbf{x}
\mapsto
\mathbf{z}_{K}(\mathbf{x}).
\end{equation}
The truncation order $K$ controls the trade-off between the dimensionality and expressiveness of the projected representation on the one hand, and its measurement and computational costs on the other. If
\begin{equation}
N_{K}
>
N_{\mathrm{F}},
\end{equation}
the mapping performs a quantum feature expansion, potentially embedding the data into a richer space. Conversely, if
\begin{equation}
N_{K}
<
N_{\mathrm{F}},
\end{equation}
it provides quantum feature compression, or quantum dimensionality reduction. Equality corresponds to a nonlinear transformation at unchanged dimensionality. The value of $K$ should therefore be regarded as a model hyperparameter rather than being constrained a priori to produce either expansion or compression. As far as the dataset considered in this work is concerned, we found better results with $N_{K}>N_{\mathrm{F}}$ (see Section~\ref{sec:results-statevector} for details).

Within this framework, it is possible to prepare a quantum circuit for encoding the original feature vector $\mathbf{x}$ into the quantum state $\ket{\psi_{\mathrm{E}}(\mathbf{x})}$, namely $\ket{\psi_{\mathrm{E}}(\mathbf{x})} = U_{\mathrm{E}}(\mathbf{x})\ket{0}^{\otimes n}$ (see Section~\ref{sec:encoding} for details). Next, it is possible to define an observable $O_K(\bm{\theta})$ given by Eq. (\ref{eqn:PQF-observable}) with maximum Pauli weight $K$ retained in the projected representation. Finally, it is possible to compute the quantum score given by Eq. (\ref{eqn:PQF-quantum-score}) as $s_{K}(\mathbf{x})=b_K+\mathbf{w}_{K}^{\mathsf{T}}\mathbf{z}_{K}(\mathbf{x})$. In this way, the quantum circuit acts as a feature extractor that maps each classical input vector $\mathbf{x}$ onto a set of projected quantum features $\mathbf{z}_{K}(\mathbf{x})$, defined as expectation values of selected single- and multi-qubit Pauli observables up to weight $K$. The resulting classical feature vectors are then provided to a conventional classifier. If all Pauli strings up to weight $K$ are retained, their number is given by Eq.~(\ref{eqn:NPK}); alternatively, structured subsets, such as nearest-neighbor observables, may be selected to further reduce the measurement cost.

For example, when binary logistic regression is used as the classical readout, the transformed classification model can be written as
\begin{equation}
\label{eqn:mapping-binary-transformed}
f_{\mathrm{LK}}:
\mathbb{R}^{N_{K}}
\to
[0,1],
\qquad
\mathbf{z}_{K}(\mathbf{x})
\mapsto
\eta_{\,\mathrm{L}}\!\left[
b_K
+
\mathbf{w}_K^{\mathsf{T}}
\mathbf{z}_{K}(\mathbf{x})
\right],
\end{equation}
where $\eta_{\,\mathrm{L}}(s)$ is given by Eq. (\ref{eqn:sigmoid}). It is possible to collect the trainable parameters of the classical model in the vector $\bm{\theta}_K=\left(b_K,\mathbf{w}_K\right)\in\mathbb{R}^{N_{\bm{\theta}}^{(K)}}$, where $N_{\bm{\theta}}=N_{\bm{\theta}}^{(K)}=N_{K}+1$. In this case, the quantum circuit determines the nonlinear feature representation, whereas the logistic-regression layer learns a linear decision boundary in the projected feature space.  Once the input features have been transformed into projected quantum features, the resulting representation can also be processed by another classical classifier, such as the previously introduced XGBoost model~\cite{Chen2016}. 

The PQF model employs a quantum circuit as a fixed feature extractor, while the subsequent classification is performed entirely by a classical model. A complementary strategy consists in making the quantum circuit itself trainable, such that both the quantum feature transformation and the measurement-based decision function can be adapted to the classification task. This approach leads naturally to variational quantum classifiers (VQCs), which are discussed in the following section.

%FFFFFFFFFFFFFFFFFFFFFFFFFFFFFFFFFFFFFFFFFF
\begin{figure}[htbp]
    \centering
    \includegraphics[scale=0.40]{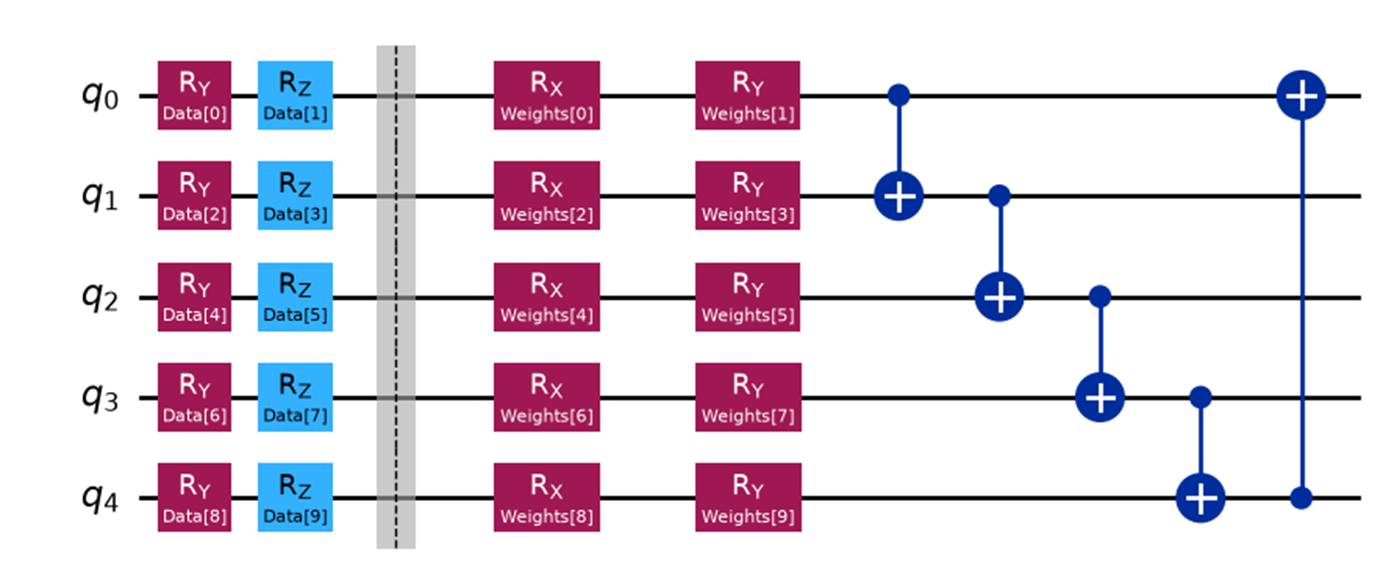}
    \caption{Example of an encoding \& trainable circuit for a Variational Quantum Classifier (VQC). The feature map $U_{\varphi\phi}(\mathbf{x})$ implements dense-angle encoding, as described in Section~\ref{sec:encoding} and, in particular, in Eq.~(\ref{eqn:dense-angle-encoding}). The trainable circuit consists of one layer ($N_l=1$) of $V_{\mathrm{XYCX}}^{\mathrm{ring}}(\bm{\theta})$, defined in Eq.~(\ref{eqn:XYCX-circular-ansatz}). In this example, $N_{\mathrm{F}}'=10$ and $N_{\bm{\theta}}=N_{\mathrm{A}}=10$.}
    \label{fig:VQC-DA-XYCX}
\end{figure}
%FFFFFFFFFFFFFFFFFFFFFFFFFFFFFFFFFFFFFFFFFF

%FFFFFFFFFFFFFFFFFFFFFFFFFFFFFFFFFFFFFFFFFF
\begin{figure}[htbp]
    \centering
    \includegraphics[scale=0.50]{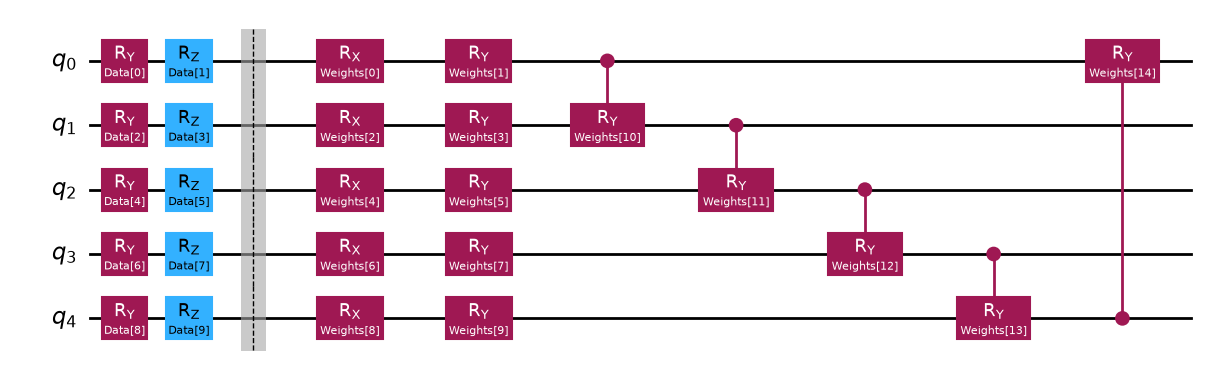}
    \caption{Example of an encoding \& trainable circuit for a Variational Quantum Classifier (VQC). The feature map $U_{\varphi\phi}(\mathbf{x})$ implements dense-angle encoding, as described in Section~\ref{sec:encoding} and, in particular, in Eq.~(\ref{eqn:dense-angle-encoding}). The trainable circuit consists of one layer ($N_l=1$) of $V_{\mathrm{XYCRy}}^{\mathrm{ring}}(\bm{\theta})$, defined in Eq.~(\ref{eqn:XYCRy-circular-ansatz}). In this example, $N_{\mathrm{F}}'=10$ and $N_{\bm{\theta}}=N_{\mathrm{A}}=15$.}
    \label{fig:VQC-DA-XYCRy}
\end{figure}
%FFFFFFFFFFFFFFFFFFFFFFFFFFFFFFFFFFFFFFFFFF

\section{Variational quantum classifier (VQC)}
\label{sec:VQC}

A standard Variational Quantum Classifier (VQC) is a hybrid quantum–classical model composed of a quantum feature map, a parameterized variational ansatz, and a measurement stage. Given a classical feature vector $\mathbf{x}$, the feature map $U_{\mathrm{E}}(\mathbf{x})$ prepares the data-dependent quantum state $\ket{\psi_{\mathrm{E}}(\mathbf{x})}$ (see Section~\ref{sec:encoding}). Then a trainable quantum circuit $V_{\mathrm{A}}(\bm{\theta})$ transforms this state into $\ket{\psi(\mathbf{x},\bm{\theta})}$, as reported in Eq. (\ref{eqn:VQC-state}). The quantum score $s_{\mathrm{V}}(\mathbf{x})$ is given by Eq. (\ref{eqn:QVC-quantum-score}) as the expectation value of a selected observable $O$, where $O=Z^{\otimes n}$. An alternative formulation consists in using the observable $O_0'=I^{\otimes (n-1)}\otimes Z\equiv Z_0$, where $Z_0$ denotes the Pauli-$Z$ operator with eigenvalues $\pm1$ acting on qubit $0$, which is the least significant qubit (LSQb). In both cases, for binary classification, this expectation value can be converted into a class probability by using Eq. (\ref{eqn:QVC-binary}). A classical optimizer iteratively updates $\bm{\theta}$ by minimizing a loss function, such as the binary cross-entropy of Eq. (\ref{eqn:loss-binary}), evaluated over the training data. The optimized quantum circuit therefore implements a nonlinear decision function whose expressivity depends on the data-encoding strategy, ansatz architecture, entanglement pattern, circuit depth, and selected observable.

The parameter-dependent unitary operator $V_{\mathrm{A}}(\bm{\theta})$ may correspond to any of the trainable circuits introduced in the following, i.e. 
\begin{equation}
\label{eqn:type-ansatz}
V_{\mathrm{A}}(\bm{\theta})\in\{V_{\mathrm{YCX}}^{\mathrm{ring}},\,V_{\mathrm{XYCX}}^{\mathrm{ring}},\,V_{\mathrm{YCRy}}^{\mathrm{ring}},\,V_{\mathrm{XYCRy}}^{\mathrm{ring}}\}.
\end{equation}

The first option consists of a layer of trainable $R_y$ rotations (see Appendix~\ref{sec:rotations}) followed by a circular (ring) arrangement of controlled-$X$ gates, namely
\begin{equation}
\label{eqn:YCX-circular-ansatz}
V_{\mathrm{YCX}}^{\mathrm{ring}}(\bm{\theta})
=
\left[
\prod_{q=0}^{n-1}
\mathrm{CX}^{\left(q,\,(q+1)\bmod n\right)}
\right]
\left[
\bigotimes_{q=0}^{n-1}
R_y^{(q)}
\!\left(\theta_q\right)
\right].
\end{equation}
In Eq.~(\ref{eqn:YCX-circular-ansatz}), the controlled-$X$ gates~\cite{nielsen_quantum_2010} are applied in ascending order of the index $q$, with $q$ denoting the control qubit and $(q+1)\bmod n$ the target qubit. For this layer altogether, the number of tunable parameters is $N_{\mathrm{YCX}}=n$. 

The second option is a more expressive gate, made of a layer of trainable $R_x$ rotations (see Appendix~\ref{sec:rotations}), followed by a second layer of trainable $R_y$ rotations, followed again by a circular arrangement of controlled-$X$ gates, namely
\begin{eqnarray} 
\label{eqn:XYCX-circular-ansatz}
V_{\mathrm{XYCX}}^{\mathrm{ring}}(\bm{\theta})
=&&
\left[
\prod_{q=0}^{n-1}
\mathrm{CX}^{\left(q,\,(q+1)\bmod n\right)}
\right]
\nonumber\\
&&\times
\left[
\bigotimes_{q=0}^{n-1}
R_y^{(q)}
\!\left(\theta_{2q+1}\right)
R_x^{(q)}
\!\left(\theta_{2q}\right)
\right].
\end{eqnarray}
For this layer altogether, the number of tunable parameters is $N_{\mathrm{XYCX}}=2\,n$. An example circuit using this trainable ansatz is depicted in Fig.~\ref{fig:VQC-DA-XYCX}.

The third option is a layer of trainable $R_y$ rotations followed by a circular arrangement of tunable controlled-$R_y$ gates, namely
\begin{eqnarray} 
\label{eqn:YCRy-circular-ansatz}
V_{\mathrm{YCRy}}^{\mathrm{ring}}(\bm{\theta})
=&&
\left[
\prod_{q=0}^{n-1}
\mathrm{CR_y}^{\left(q,\,(q+1)\bmod n\right)}
\!\left(\theta_{q+n}\right)
\right]
\nonumber\\
&&\times
\left[
\bigotimes_{q=0}^{n-1}
R_y^{(q)}
\!\left(\theta_q\right)
\right].
\end{eqnarray}

Finally, the fourth option consists in substituting a circular arrangement of tunable controlled-$R_y$ gates to the circular arrangement of controlled-$X$ gates in the previous circuit given by Eq. (\ref{eqn:XYCX-circular-ansatz}), namely
\begin{eqnarray} 
\label{eqn:XYCRy-circular-ansatz}
V_{\mathrm{XYCRy}}^{\mathrm{ring}}(\bm{\theta})
=&&
\left[
\prod_{q=0}^{n-1}
\mathrm{CR_y}^{\left(q,\,(q+1)\bmod n\right)}
\!\left(\theta_{q+2n}\right)
\right]
\nonumber\\
&&\times
\left[
\bigotimes_{q=0}^{n-1}
R_y^{(q)}
\!\left(\theta_{2q+1}\right)
R_x^{(q)}
\!\left(\theta_{2q}\right)
\right].
\end{eqnarray}
An example circuit using this trainable ansatz is depicted in Fig.~\ref{fig:VQC-DA-XYCRy}.

In both Eq.~(\ref{eqn:YCRy-circular-ansatz}) and Eq.~(\ref{eqn:XYCRy-circular-ansatz}), the controlled rotations are applied in ascending order of $q$, with $q$ and $(q+1)\bmod n$ denoting the control and target qubits, respectively. These ansatz circuits therefore contain $2n$ and $3n$ trainable parameters, namely $N_{\mathrm{YCRy}}=2\,n$ and $N_{\mathrm{XYCRy}}=3\,n$, respectively. For the null parameter vector $\bm{\theta}=\bm{0}$, all rotations reduce to the identity, so that $V_{\mathrm{YCRy}}^{\mathrm{ring}}(\bm{0})=V_{\mathrm{XYCRy}}^{\mathrm{ring}}(\bm{0})=I^{\otimes n}$; hence, the ansatz leaves the quantum state unchanged. 

The preceding gate sequence can be repeated over $N_l$ consecutive layers to increase the expressive power of the trainable circuit. However, increasing the circuit depth may impair trainability, for example by promoting the emergence of barren plateaus~\cite{nielsen_quantum_2010}. In general, given $N_{l}$, the number of tunable parameters is $N_{\bm{\theta}}=N_{\bm{\theta}}^{\mathrm{(V)}}=N_{\mathrm{A}}\times N_{l}$.

\section{Logistic-regression-anchored (LRA) VQC}
\label{sec:LRA-VQC}

\subsection{LRA by cos-angle encoding}
\label{sec:LRA-VQC-cos-angle}

A conventional VQC is generally initialized from randomly selected circuit parameters and trained without explicitly exploiting the solution provided by a classical classifier. This may lead to a difficult optimization problem and, in particular for small tabular datasets, to a deterioration of the generalization performance relative to simpler classical models. To address this issue, we introduce a logistic-regression-anchored variational quantum classifier (LRA), in which a previously trained logistic-regression model (see Section~\ref{sec:logistic}) is embedded exactly into the quantum model and subsequently enriched through a trainable variational circuit. The key idea behind embedding logistic regression into a quantum model relies on two ingredients: (i) encoding the input features such that each feature can be easily extracted by simple measurements, and (ii) constructing the final readout observable as a linear combination of operators, with coefficients determined by the parameters of a pretrained logistic-regression model.

Under angle encoding (see Section~\ref{sec:angle} for details), one qubit is assigned to each input feature, so that $n=N_{\mathrm{F}}$. Let the input features be rescaled to the interval $[-1,1]$ as in Eq. (\ref{eqn:angle-scaling2}), such that $\mathbf{x}' \in[-1,1]^{N_F}$. The classical input is encoded through the product state without entanglement given by Eq. (\ref{eqn:angle-encoding2}), namely $\ket{\psi_{\mathcal{E}}(\mathbf{x})}=\ket{\psi_{\varphi'}(\mathbf{x})}=U_{\varphi'}(\mathbf{x})
\ket{0}^{\otimes n}$, because the cos-angle encoding ensures a simple extraction of the features. Indeed, this encoding satisfies
\begin{equation}
\label{eqn:LRA-local-expectation}
\bra{\psi_{\varphi'}(\mathbf{x})}
Z_q
\ket{\psi_{\varphi'}(\mathbf{x})}
=x'_q,
\end{equation}
where $Z_q$ denotes the Pauli-$Z$ operator acting on qubit $q$. Consider a logistic-regression model trained on the same rescaled features,
\begin{equation}
\label{eqn:LRA-classical-logit}
s_{\mathrm{L}}'(\mathbf{x})
=
b_{\mathrm{L}}
+
\mathbf{w}_{\mathrm{L}}^{\mathsf{T}}\mathbf{x}'
=
b_{\mathrm{L}}
+
\sum_{q=0}^{n-1}
w_{\mathrm{L},q}x_q',
\end{equation}
with the probability that the sample is a thermoelectric material equal to 
\begin{equation}
\label{eqn:LRA-classical-probability}
p'_{\mathrm{L}}(y=1\mid\mathbf{x})
=
\eta_{\,\mathrm{L}}\!\left(
s'_{\mathrm{L}}(\mathbf{x})
\right),
\end{equation}
where $\eta_{\,\mathrm{L}}(s)$ is the sigmoid function given by Eq. (\ref{eqn:sigmoid}). The corresponding coefficients can be incorporated into the trainable-observable readout
\begin{equation}
\label{eqn:LRA-observable}
O_{\varphi'}(\bm{\theta}_{\mathrm{L}})
=
b_{\mathrm{L}} I^{\otimes n}
+
\sum_{q=0}^{n-1}
w_{\mathrm{L},q}Z_q.
\end{equation}
Using Eq.~(\ref{eqn:LRA-local-expectation}), its expectation value becomes
\begin{eqnarray} 
\label{eqn:anchor}
\braket{\psi_{\varphi'}(\mathbf{x})
|O_{\varphi'}(\bm{\theta}_{\mathrm{L}})|
\psi_{\varphi'}(\mathbf{x})}
&=&
b_{\mathrm{L}}
+
\sum_{q=0}^{n-1}
w_{\mathrm{L},q}
\bra{\psi'(\mathbf{x})}
Z_q
\ket{\psi'(\mathbf{x})}
\nonumber\\
&=&
b_{\mathrm{L}}
+
\sum_{q=0}^{n-1}
w_{\mathrm{L},q}x_q'
=
s_{\mathrm{L}}'(\mathbf{x}),
\end{eqnarray}
which substantially coincides with Eq. (\ref{eqn:quantum-logit-LRA-observable}), as far as the different scaling is neglected. Therefore, before introducing any variational correction, the quantum model reproduces exactly the logit and probability predicted by the reference logistic-regression classifier.

Among the parameter-dependent unitary operators discussed in Section~\ref{sec:VQC}, the classical baseline is then enriched by applying the parameterized quantum circuit $V_{\mathcal{A}}(\bm{\theta})=V_{\mathrm{YCRy}}^{\mathrm{ring}}(\bm{\theta})$ given by Eq. (\ref{eqn:YCRy-circular-ansatz}), after the encoding:
\begin{equation}
\label{eqn:LRA-state}
\ket{\psi(\mathbf{x},{\bm{\theta}})}
=
V_{\mathrm{YCRy}}^{\mathrm{ring}}(\bm{\theta})\,
U_{\varphi'}(\mathbf{x})
\ket{0}^{\otimes n}.
\end{equation}
In the present implementation, the variational correction consists of a layer of trainable $R_y$ rotations with local tunable parameters, followed by a circular arrangement of tunable controlled-$R_y$ gates with interaction parameters. In Eq. (\ref{eqn:YCRy-circular-ansatz}), the tensor product is used for the local $R_y$ layer because its gates act on distinct qubits, whereas the controlled-$R_y$ gates acting on qubit pairs must be composed through an operator product. This sequence can be repeated over $N_l$ consecutive
layers to increase the expressive power of the trainable circuit, at the risk of promoting the emergence of barren plateaus
\cite{nielsen_quantum_2010}. In general, given $N_{l}$, the number of tunable parameters is $N_{\bm{\theta}}=N_{\bm{\theta}}^{\mathrm{(A)}}=N_{\mathrm{YCRy}}\times N_{l}=2\,n\,N_{l}$. In the present context, the most important feature of this ansatz is that $V_{\mathrm{YCRy}}^{\mathrm{ring}}(\bm{0})=I^{\otimes n}$, which ensures the proper anchoring to the previous logistic regression model given by Eq. (\ref{eqn:anchor}).

The LRA logit and the corresponding probability of the sample being a thermoelectric material are defined as
\begin{equation}
\label{eqn:LRA-quantum-logit}
s'_{\mathrm{A}}(\mathbf{x})=
\bra{\psi(\mathbf{x},{\bm{\theta}})}
O_{\varphi'}(\bm{\theta}_{\mathrm{L}})
\ket{\psi(\mathbf{x},{\bm{\theta}})},
\end{equation}
where $p'_{\mathrm{A}}(y=1\mid\mathbf{x}) = \eta_{\,\mathrm{L}}\!\left( s'_{\mathrm{A}}(\mathbf{x})\right)$ and $\eta_{\,\mathrm{L}}(s)$ is the sigmoid function given by Eq. (\ref{eqn:sigmoid}). At initialization, $s'_{\mathrm{A}}(\mathbf{x})=s'_{\mathrm{L}}(\mathbf{x})$ and hence $p'_{\mathrm{A}}(\mathbf{x})=p'_{\mathrm{L}}(\mathbf{x})$, whereas nonzero values of the parameters introduce trainable nonlinear and inter-feature corrections to the original linear decision function.

The variational parameters can be determined by minimizing a regularized binary cross-entropy objective similar to Eq. (\ref{eqn:loss-binary}), usually with $\lambda_j=\lambda$ for all $0\leq j\leq N_{\bm{\theta}}-1$. The parameter $\lambda\geq0$ is an anchoring penalty that discourages unnecessarily large departures from the logistic-regression solution. In the limit of a large $\lambda$, the model remains close to its classical baseline, whereas smaller values allow increasingly flexible quantum corrections. In the present work, for the sake of simplicity, we used $\lambda=0$. The central feature of the proposed construction is therefore the nesting of the reference classical model within the variational quantum model. Unlike a randomly initialized VQC, the LRA starts from a meaningful and reproducible decision function, while the identity-initialized ansatz provides additional degrees of freedom that may capture nonlinear dependencies and interactions among the input features. Nevertheless, any improvement in predictive performance remains data dependent and must be assessed on validation and test sets.

For clarity, we illustrate the construction using the angle-encoding scheme introduced in Section~\ref{sec:encoding}. On the other hand, the next section discusses the implementation of the proposed approach in case of probability encoding preserving the size.

\subsection{LRA by probability encoding with size}
\label{sec:LRA-VQC-prob-size}

The cos-angle encoding discussed in the previous section may require a large number of qubits in case of a large number of features. In this section, we use instead the probability encoding with size (see Section~\ref{sec:amplitude} for details), which implies $n
=\left\lceil\log_2\!\left(N_{\mathrm{F}}+1\right)\right\rceil$. 
Let the input features be rescaled to the interval $[0,1]$ as in Eq. (\ref{eqn:angle-scaling}), such that $\mathbf{x}\in[0,1]^{N_F}$. The classical input is encoded through the gate given by Eq. (\ref{eqn:auxiliary-state-encoding}), namely $\ket{\psi_{\mathcal{E}}(\mathbf{x})}=\ket{\psi_{\mathrm{A}''}(\mathbf{x})}=U_{\mathrm{A}''}(\mathbf{x})\ket{0}^{\otimes n}$, because this encoding ensures a simple extraction of the features. Indeed, this encoding satisfies
\begin{equation}
\label{eqn:LRA-P-local-expectation}
\bra{\psi_{\mathrm{A}''}(\mathbf{x})}
\ket{\nu}\bra{\nu}
\ket{\psi_{\mathrm{A}''}(\mathbf{x})}
=\frac{x_\nu}{S_{\max}},
\end{equation}
for $0\leq \nu \leq N_{\mathrm{F}}-1$. Consider a logistic-regression model trained on the same rescaled features given by Eq. (\ref{eqn:logit}), with the probability that the sample is a thermoelectric material given by Eq. (\ref{eqn:binary}). The corresponding coefficients can be incorporated into the trainable-observable readout
\begin{equation}
\label{eqn:LRA-P-observable}
O_{\mathrm{A}''}(\bm{\theta}_{\mathrm{L}})
=
b_{\mathrm{L}} I^{\otimes n}
+S_{\max}
\sum_{\nu=0}^{N_{\mathrm{F}}-1}
w_{\mathrm{L},\nu}\ket{\nu}\bra{\nu}.
\end{equation}
Using Eq. (\ref{eqn:LRA-P-local-expectation}), its expectation value becomes
\begin{equation}
\label{eqn:P-anchor}
\braket{\psi_{\mathrm{A}''}(\mathbf{x})
|O_{\mathrm{A}''}(\bm{\theta}_{\mathrm{L}})|
\psi_{\mathrm{A}''}(\mathbf{x})}
=b_{\mathrm{L}}+
\mathbf{w}_{\mathrm{L}}^{\mathsf{T}}\mathbf{x}
=s_{\mathrm{L}}(\mathbf{x}),
\end{equation}
which coincides with Eq. (\ref{eqn:quantum-logit-LRA-observable}). Again, before introducing any variational correction, the
quantum model reproduces exactly the logit and probability
predicted by the reference logistic-regression classifier.

Among the parameter-dependent unitary operators discussed in Section~\ref{sec:VQC}, the classical baseline is then enriched by applying again the parameterized quantum circuit $V_{\mathcal{A}}(\bm{\theta})=V_{\mathrm{YCRy}}^{\mathrm{ring}}(\bm{\theta})$ given by Eq. (\ref{eqn:YCRy-circular-ansatz}), as done in the previous section. In the present context, the most important feature of this ansatz is that $V_{\mathrm{YCRy}}^{\mathrm{ring}}(\bm{0})=I^{\otimes n}$, which ensures the proper anchoring to the previous logistic regression model given by Eq. (\ref{eqn:P-anchor}). In case of probability encoding with size, the LRA logit is defined as
\begin{equation}
\label{eqn:LRA-P-quantum-logit}
s_{\mathrm{A}}(\mathbf{x})=
\bra{\psi(\mathbf{x},{\bm{\theta}})}
O_{\mathrm{A}''}(\bm{\theta}_{\mathrm{L}})
\ket{\psi(\mathbf{x},{\bm{\theta}})},
\end{equation}
and the corresponding probability of the sample being a thermoelectric material as $p_{\mathrm{A}}(y=1\mid\mathbf{x}) = \eta_{\,\mathrm{L}}\!\left( s_{\mathrm{A}}(\mathbf{x})\right)$, where $\eta_{\,\mathrm{L}}(s)$ is the sigmoid function given by Eq. (\ref{eqn:sigmoid}). At initialization, $s_{\mathrm{A}}(\mathbf{x})=s_{\mathrm{L}}(\mathbf{x})$ and hence $p_{\mathrm{A}}(\mathbf{x})=p_{\mathrm{L}}(\mathbf{x})$, whereas nonzero values of the parameters introduce trainable nonlinear and inter-feature corrections to the original linear decision function.

%TTTTTTTTTTTTTTTTTTTTTTTTTTTTTTTTTTTTTTTTTT
\begin{table*}[!t]
\caption{Predictive performance of the classical and VQC classification models on the training and validation sets. As discussed in Section~\ref{sec:egnome}, to investigate model performance in a data-limited training regime, $N_{\mathrm{train}}=700$ and $N_{\mathrm{valid}}=300$ samples were assigned to the training and validation sets, respectively. The original dataset comprises $N_{\mathrm{F}}=145$ features, whereas the $N_{\mathrm{F}}'=10$ most informative features were selected to facilitate future implementations on actual quantum hardware. The VQC models are described in Section~\ref{sec:VQC}. The reported metrics are binary cross-entropy (BCE), area under the receiver operating characteristic curve (ROC--AUC), area under the precision--recall curve (PR--AUC), precision (P), recall (R), and $F_1$-score. During VQC optimization, training was terminated when the change in BCE satisfied $|\Delta\mathrm{BCE}|<10^{-5}$ for three consecutive checks, performed every ten objective-function evaluations.}
\label{tab:new-VQC-table}

\centering
\scriptsize
\setlength{\tabcolsep}{2.5pt}
\renewcommand{\arraystretch}{1.15}

\begin{tabularx}{\textwidth}{
@{}rllrlr
*{9}{X}
@{}
}
\toprule
\multicolumn{6}{c}{}
&
\multicolumn{3}{c}{Training}
&
\multicolumn{6}{c}{Validation}
\\
\cmidrule(lr){7-9}
\cmidrule(lr){10-15}

$N_{\mathrm{F}}$
& Model
& Encoding
& $n$
& Ansatz
& $N_{\bm{\theta}}$
& BCE
& \shortstack{ROC--\\AUC}
& \shortstack{PR--\\AUC}
& BCE
& \shortstack{ROC--\\AUC}
& \shortstack{PR--\\AUC}
& P
& R
& \shortstack{$F_1$-\\score}
\\
\midrule

145 & XGBoost & --- & --- & --- & --- & 0.0789 & 0.9995 & 0.9995 & 0.1984 & 0.9788 & 0.9789 & 0.9045 & 0.9467 & 0.9251 \\
\midrule
145 & VQC.01 & Amplitude & 8 & $\mathrm{YCX}\times 3$ & 24 & 0.6095 & 0.8756 & 0.8875 & 0.6174 & 0.8449 & 0.8630 & 0.7159 & 0.8400 & 0.7730 \\
145 & \textbf{VQC.02} & \textbf{Amplitude} & 8 & $\bm{\mathrm{YCX}\times 4}$ & 32 & 0.5799 & 0.8970 & 0.9052 & 0.5858 & 0.8730 & 0.8828 & 0.7430 & 0.8867 & 0.8085 \\
145 & VQC.03 & Amplitude & 8 & $\mathrm{YCRy}\times 3$ & 48 & 0.5640 & 0.8974 & 0.9048 & 0.5759 & 0.8628 & 0.8636 & 0.7314 & 0.8533 & 0.7877 \\
145 & VQC.04 & Amplitude & 8 & $\mathrm{YCRy}\times 4$ & 64 & 0.5525 & 0.8996 & 0.9027 & 0.5665 & 0.8626 & 0.8706 & 0.7267 & 0.8333 & 0.7764 \\
145 & VQC.05 & Amplitude & 8 & $\mathrm{XYCX}\times 3$ & 48 & 0.5861 & 0.8795 & 0.8814 & 0.5987 & 0.8416 & 0.8472 & 0.6919 & 0.8533 & 0.7642 \\
145 & VQC.06 & Amplitude & 8 & $\mathrm{XYCX}\times 4$ & 64 & 0.5767 & 0.8854 & 0.8846 & 0.5892 & 0.8410 & 0.8463 & 0.6935 & 0.8600 & 0.7679 \\
145 & VQC.07 & Amplitude & 8 & $\mathrm{XYCRy}\times 3$ & 72 & 0.5605 & 0.8895 & 0.9006 & 0.5691 & 0.8600 & 0.8691 & 0.7278 & 0.8200 & 0.7712 \\
145 & VQC.08 & Amplitude & 8 & $\mathrm{XYCRy}\times 4$ & 96 & 0.5495 & 0.8984 & 0.9053 & 0.5637 & 0.8627 & 0.8695 & 0.7514 & 0.8667 & 0.8050 \\
%
%\midrule
%
%145 & \textcolor{red}{VQC-TE} & \textcolor{red}{Trainable Encoding} & 8 & $\textcolor{red}{\mathrm{YZ}_{\mathrm{TE}}+\mathrm{CRy}}$ & 42 & \multicolumn{3}{c}{\textcolor{red}{first proposal by ChatGPT}} & 0.5837 & 0.7636 & \textcolor{red}{0.8055} & 0.8120 & 0.6333 & 0.7116 \\
%
\midrule
10 & XGBoost & --- & --- & --- & --- & 0.1077 & 0.9976 & 0.9977 & 0.2156 & 0.9749 & 0.9743 & 0.8854 & 0.9267 & 0.9055 \\
\midrule
10 & VQC.09 & Angle & 10 & $\mathrm{YCX}\times 3$ & 30 & 0.3949 & 0.9269 & 0.9384 & 0.4500 & 0.8898 & 0.8955 & 0.7529 & 0.8733 & 0.8086 \\
10 & VQC.10 & Angle & 10 & $\mathrm{YCX}\times 4$ & 40 & 0.4225 & 0.9160 & 0.9331 & 0.4806 & 0.8699 & 0.8662 & 0.7143 & 0.8667 & 0.7831 \\
10 & VQC.11 & Angle & 10 & $\mathrm{YCRy}\times 3$ & 60 & 0.4792 & 0.9192 & 0.9286 & 0.5124 & 0.8805 & 0.8764 & 0.7433 & 0.9267 & 0.8249 \\
10 & VQC.12 & Angle & 10 & $\mathrm{YCRy}\times 4$ & 80 & 0.3822 & 0.9175 & 0.9285 & 0.4509 & 0.8786 & 0.8901 & 0.7593 & 0.8200 & 0.7885 \\
10 & VQC.13 & Angle & 10 & $\mathrm{XYCX}\times 3$ & 60 & 0.3538 & 0.9246 & 0.9325 & 0.4407 & 0.8772 & 0.8910 & 0.7532 & 0.7733 & 0.7632 \\
10 & VQC.14 & Angle & 10 & $\mathrm{XYCX}\times 4$ & 80 & 0.3872 & 0.9139 & 0.9257 & 0.4530 & 0.8780 & 0.8884 & 0.7605 & 0.8467 & 0.8013 \\
10 & VQC.15 & Angle & 10 & $\mathrm{XYCRy}\times 3$ & 90 & 0.3654 & 0.9365 & 0.9492 & 0.4161 & 0.9065 & 0.9155 & 0.8000 & 0.9067 & 0.8500 \\
10 & VQC.16 & Angle & 10 & $\mathrm{XYCRy}\times 4$ & 120 & 0.3459 & 0.9317 & 0.9422 & 0.4280 & 0.8850 & 0.8922 & 0.7818 & 0.8600 & 0.8190 \\
\midrule
10 & VQC.17 & Dense angle & 5 & $\mathrm{YCX}\times 3$ & 15 & 0.3671 & 0.9207 & 0.9227 & 0.4567 & 0.8774 & 0.8754 & 0.7554 & 0.9267 & 0.8323 \\
10 & VQC.18 & Dense angle & 5 & $\mathrm{YCX}\times 4$ & 20 & 0.4095 & 0.9139 & 0.9259 & 0.4635 & 0.8675 & 0.8696 & 0.7515 & 0.8267 & 0.7873 \\
10 & VQC.19 & Dense angle & 5 & $\mathrm{YCRy}\times 3$ & 30 & 0.3997 & 0.9149 & 0.9278 & 0.4660 & 0.8625 & 0.8827 & 0.8231 & 0.8067 & 0.8148 \\
10 & VQC.20 & Dense angle & 5 & $\mathrm{YCRy}\times 4$ & 40 & 0.3427 & 0.9346 & 0.9430 & 0.4299 & 0.8926 & 0.9013 & 0.7654 & 0.9133 & 0.8328 \\
10 & VQC.21 & Dense angle & 5 & $\mathrm{XYCX}\times 3$ & 30 & 0.3782 & 0.9343 & 0.9413 & 0.4280 & 0.8952 & 0.8970 & 0.7683 & 0.8400 & 0.8025 \\
10 & \textbf{VQC.22} & \textbf{Dense angle} & 5 & $\bm{\mathrm{XYCX}\times 4}$ & 40 & 0.3289 & 0.9411 & 0.9482 & 0.3976 & 0.9070 & 0.9091 & 0.8012 & 0.9133 & 0.8536 \\
10 & VQC.23 & Dense angle & 5 & $\mathrm{XYCRy}\times 3$ & 45 & 0.3221 & 0.9435 & 0.9495 & 0.4116 & 0.8982 & 0.9016 & 0.7898 & 0.9267 & 0.8528 \\
10 & VQC.24 & Dense angle & 5 & $\mathrm{XYCRy}\times 4$ & 60 & 0.3195 & 0.9476 & 0.9530 & 0.3993 & 0.9070 & 0.9102 & 0.7977 & 0.9200 & 0.8545 \\
\midrule
10 & VQC.25 & Amplitude & 4 & $\mathrm{YCX}\times 3$ & 12 & 0.4926 & 0.9028 & 0.9001 & 0.5338 & 0.8468 & 0.8179 & 0.7333 & 0.8800 & 0.8000 \\
10 & VQC.26 & Amplitude & 4 & $\mathrm{YCX}\times 4$ & 16 & 0.4549 & 0.9177 & 0.9240 & 0.4979 & 0.8737 & 0.8848 & 0.7625 & 0.8133 & 0.7871 \\
10 & VQC.27 & Amplitude & 4 & $\mathrm{YCRy}\times 3$ & 24 & 0.4481 & 0.9197 & 0.9290 & 0.4938 & 0.8748 & 0.8837 & 0.7834 & 0.8200 & 0.8013 \\
10 & VQC.28 & Amplitude & 4 & $\mathrm{YCRy}\times 4$ & 32 & 0.4504 & 0.9177 & 0.9247 & 0.4959 & 0.8722 & 0.8768 & 0.7799 & 0.8267 & 0.8026 \\
10 & VQC.29 & Amplitude & 4 & $\mathrm{XYCX}\times 3$ & 24 & 0.4594 & 0.9184 & 0.9184 & 0.5054 & 0.8717 & 0.8583 & 0.7557 & 0.8867 & 0.8160 \\
10 & VQC.30 & Amplitude & 4 & $\mathrm{XYCX}\times 4$ & 32 & 0.4554 & 0.9174 & 0.9241 & 0.5009 & 0.8706 & 0.8763 & 0.7771 & 0.8133 & 0.7948 \\
10 & VQC.31 & Amplitude & 4 & $\mathrm{XYCRy}\times 3$ & 36 & 0.4497 & 0.9198 & 0.9277 & 0.4936 & 0.8753 & 0.8833 & 0.7785 & 0.8200 & 0.7987 \\
10 & VQC.32 & Amplitude & 4 & $\mathrm{XYCRy}\times 4$ & 48 & 0.4483 & 0.9209 & 0.9295 & 0.4935 & 0.8773 & 0.8846 & 0.7875 & 0.8400 & 0.8129 \\
\bottomrule
\multicolumn{15}{@{}p{\linewidth}@{}}{%
    \footnotesize
    \textit{Note:} A compact notation is adopted for the ansatz circuits: $\mathrm{YCX}$, 
    $\mathrm{YCRy}$, $\mathrm{XYCX}$, and $\mathrm{XYCRy}$ denote $V_{\mathrm{YCX}}^{\mathrm{ring}}(\bm{\theta})$, $V_{\mathrm{YCRy}}^{\mathrm{ring}}(\bm{\theta})$, $V_{\mathrm{XYCX}}^{\mathrm{ring}}(\bm{\theta})$, and $V_{\mathrm{XYCRy}}^{\mathrm{ring}}(\bm{\theta})$ , respectively.
}
\end{tabularx}
\end{table*}
%TTTTTTTTTTTTTTTTTTTTTTTTTTTTTTTTTTTTTTTTTT

%TTTTTTTTTTTTTTTTTTTTTTTTTTTTTTTTTTTTTTTTTT
\begin{table*}[!t]
\caption{Predictive performance of the classical and PQF classification models on the training and validation sets. The datasets, data partitions, and reported performance metrics are the same as those described in the caption of Table~\ref{tab:new-VQC-table}. The PQF models are presented in Section~\ref{sec:projected}.}
\label{tab:new-PQF-table}

\centering
\scriptsize
\setlength{\tabcolsep}{2.5pt}
\renewcommand{\arraystretch}{1.15}

\begin{tabularx}{\textwidth}{
@{}lllrl
*{9}{X}
@{}
}
\toprule
\multicolumn{5}{c}{}
&
\multicolumn{3}{c}{Training}
&
\multicolumn{6}{c}{Validation}
\\
\cmidrule(lr){6-8}
\cmidrule(lr){9-14}

$N_F$
& Model
& Encoding
& $n$
& $N_{\bm{\theta}}$
& BCE
& \shortstack{ROC--\\AUC}
& \shortstack{PR--\\AUC}
& BCE
& \shortstack{ROC--\\AUC}
& \shortstack{PR--\\AUC}
& P
& R
& \shortstack{$F_1$-\\score}
\\
\midrule

145 & XGBoost & --- & --- & --- & 0.0789 & 0.9995 & 0.9995 & 0.1984 & 0.9788 & 0.9789 & 0.9045 & 0.9467 & 0.9251 \\
\midrule
145 & PQF.01 & Amplitude & 8 & 25 ($K=1$) & 0.2031 & 0.9896 & 0.9883 & 0.3384 & 0.9394 & 0.9443 & 0.8232 & 0.9000 & 0.8599 \\
145 & PQF.02 & Amplitude & 8 & 277 ($K=2$) & 0.1007 & 0.9993 & 0.9993 & 0.3172 & 0.9394 & 0.9297 & 0.8263 & 0.9200 & 0.8707 \\
145 & \textbf{PQF.03} & \textbf{Amplitude} & 8 & 1789 ($\bm{K=3}$) & 0.0650 & 1.0000 & 1.0000 & 0.2652 & 0.9618 & 0.9673 & 0.8704 & 0.9400 & 0.9038 \\
%
%145 & PQF.xx & Amplitude & 8 & 7459 ($K=4$) & 0.051 & 1.000 & 1.000 & 0.270 & 0.960 & 0.965 & 0.870 & 0.940 & 0.904 \\
%
\midrule
10 & XGBoost & --- & --- & --- & 0.1077 & 0.9976 & 0.9977 & 0.2156 & 0.9749 & 0.9743 & 0.8854 & 0.9267 & 0.9055 \\
\midrule
%
%10 & \textbf{PQF.04} & \textbf{Angle} & 10 & 31 ($\bm{K=1}$) & 0.094 & 0.999 & 0.999 & 0.187 & 0.982 & 0.981 & 0.910 & 0.947 & 0.928 \\
%
10 & \textbf{PQF.04} & \textbf{Angle} & 10 & 31 ($\bm{K=1}$) & 0.0941 & 0.9986 & 0.9986 & 0.1871 & 0.9819 & 0.9814 & 0.9103 & 0.9467 & 0.9281 \\
%
%10 & PQF.05 & Angle & 10 & 436 ($K=2$) & 0.067 & 1.000 & 1.000 & 0.222 & 0.971 & 0.973 & 0.884 & 0.967 & 0.924 \\
%
10 & PQF.05 & Angle & 10 & 436 ($K=2$) & 0.0673 & 0.9998 & 0.9998 & 0.2223 & 0.9714 & 0.9725 & 0.8841 & 0.9667 & 0.9236 \\
%
%10 & PQF.06 & Angle & 10 & 3676 ($K=3$) & 0.047 & 1.000 & 1.000 & 0.219 & 0.971 & 0.973 & 0.909 & 0.933 & 0.921 \\
%
10 & PQF.06 & Angle & 10 & 3676 ($K=3$) & 0.0474 & 1.0000 & 1.0000 & 0.2191 & 0.9713 & 0.9730 & 0.9091 & 0.9333 & 0.9211 \\
%
%10 & PQF.07 & Dense angle & 5 & 16 ($K=1$) & 0.122 & 0.997 & 0.997 & 0.248 & 0.965 & 0.967 & 0.901 & 0.913 & 0.907 \\
%
10 & PQF.07 & Dense angle & 5 & 16 ($K=1$) & 0.1221 & 0.9972 & 0.9974 & 0.2482 & 0.9651 & 0.9670 & 0.9013 & 0.9133 & 0.9073 \\
%
%10 & PQF.08 & Dense angle & 5 & 106 ($K=2$) & 0.092 & 0.999 & 0.999 & 0.291 & 0.949 & 0.953 & 0.874 & 0.880 & 0.877 \\
%
10 & PQF.08 & Dense angle & 5 & 106 ($K=2$) & 0.0923 & 0.9991 & 0.9992 & 0.2912 & 0.9493 & 0.9526 & 0.8742 & 0.8800 & 0.8771 \\
%
%10 & PQF.09 & Dense angle & 5 & 376 ($K=3$) & 0.075 & 1.000 & 1.000 & 0.287 & 0.950 & 0.955 & 0.871 & 0.900 & 0.885 \\
%
10 & PQF.09 & Dense angle & 5 & 376 ($K=3$) & 0.0748 & 0.9998 & 0.9998 & 0.2867 & 0.9498 & 0.9554 & 0.8710 & 0.9000 & 0.8852 \\
%
%10 & PQF.10 & Amplitude & 4 & 13 ($K=1$) & 0.186 & 0.986 & 0.987 & 0.368 & 0.915 & 0.903 & 0.847 & 0.847 & 0.847 \\
%
10 & PQF.10 & Amplitude & 4 & 13 ($K=1$) & 0.1855 & 0.9862 & 0.9865 & 0.3682 & 0.9149 & 0.9032 & 0.8467 & 0.8467 & 0.8467 \\
%
%10 & PQF.11 & Amplitude & 4 & 67 ($K=2$) & 0.129 & 0.997 & 0.997 & 0.361 & 0.926 & 0.931 & 0.830 & 0.847 & 0.838 \\
%
10 & PQF.11 & Amplitude & 4 & 67 ($K=2$) & 0.1292 & 0.9967 & 0.9968 & 0.3612 & 0.9260 & 0.9310 & 0.8301 & 0.8467 & 0.8383 \\
%
%10 & PQF.12 & Amplitude & 4 & 175 ($K=3$) & 0.107 & 0.999 & 0.999 & 0.298 & 0.950 & 0.953 & 0.864 & 0.887 & 0.875 \\
%
10 & PQF.12 & Amplitude & 4 & 175 ($K=3$) & 0.1070 & 0.9987 & 0.9987 & 0.2976 & 0.9496 & 0.9532 & 0.8636 & 0.8867 & 0.8750 \\
\bottomrule
% \multicolumn{15}{@{}p{\linewidth}@{}}{%
%     \footnotesize
%     \textit{Note:} A compact notation is adopted for the ansatz circuits:
%     $\mathrm{YCRy}$, $\mathrm{YCX}$, and $\mathrm{XYCX}$ denote
%     $V_{\mathrm{YCRy}}^{\mathrm{ring}}(\bm{\theta})$,
%     $V_{\mathrm{YCX}}^{\mathrm{ring}}(\bm{\theta})$, and
%     $V_{\mathrm{XYCX}}^{\mathrm{ring}}(\bm{\theta})$, respectively.
% }
\end{tabularx}
\end{table*}
%TTTTTTTTTTTTTTTTTTTTTTTTTTTTTTTTTTTTTTTTTT

%TTTTTTTTTTTTTTTTTTTTTTTTTTTTTTTTTTTTTTTTTT
\begin{table*}[!t]
\caption{Predictive performance of the classical and LRA-VQC classification models on the training and validation sets. The datasets, data partitions, optimization-convergence criterion, and performance metrics are the same as those described in the caption of Table~\ref{tab:new-VQC-table}. The LRA-VQC models are presented in Section~\ref{sec:LRA-VQC}, in particular, LRA by cos-angle encoding in Section~\ref{sec:LRA-VQC-cos-angle} and LRA by probability encoding with size in Section~\ref{sec:LRA-VQC-prob-size}.}
\label{tab:new-LRA-table}

\centering
\scriptsize
\setlength{\tabcolsep}{2.5pt}
\renewcommand{\arraystretch}{1.15}

\begin{tabularx}{\textwidth}{
@{}rllrlr
*{9}{X}
@{}
}
\toprule
\multicolumn{6}{c}{}
&
\multicolumn{3}{c}{Training}
&
\multicolumn{6}{c}{Validation}
\\
\cmidrule(lr){7-9}
\cmidrule(lr){10-15}

$N_{\mathrm{F}}$
& Model
& Encoding
& $n$
& Ansatz
& $N_{\bm{\theta}}$
& BCE
& \shortstack{ROC--\\AUC}
& \shortstack{PR--\\AUC}
& BCE
& \shortstack{ROC--\\AUC}
& \shortstack{PR--\\AUC}
& P
& R
& \shortstack{$F_1$-\\score}
\\
\midrule

145 & XGBoost & --- & --- & --- & --- & 0.0789 & 0.9995 & 0.9995 & 0.1984 & 0.9788 & 0.9789 & 0.9045 & 0.9467 & 0.9251 \\
\midrule
%
%145 & LRA.01 old & Prob. with size & 8 & $\mathrm{YCRy}$ & 16 & 0.2086 & 0.9764 & 0.9772 & 0.3277 & 0.9396 & 0.9459 & 0.8225 & 0.9267 & 0.8715 \\
%
145 & \textbf{LRA.01} & \textbf{Prob. with size} & 8 & $\bm{\mathrm{YCRy}}$ & 16 & 0.2083 & 0.9767 & 0.9775 & 0.3294 & 0.9393 & 0.9453 & 0.8129 & 0.9267 & 0.8660 \\
%
%145 & LRA.02 old & Prob. with size & 8 & $\mathrm{YCRy}\times 2$ & 32 & 0.2093 & 0.9764 & 0.9772 & 0.3281 & 0.9395 & 0.9458 & 0.8176 & 0.9267 & 0.8688 \\
%
145 & LRA.02 & Prob. with size & 8 & $\mathrm{YCRy}\times 2$ & 32 & 0.2079 & 0.9766 & 0.9774 & 0.3292 & 0.9394 & 0.9456 & 0.8129 & 0.9267 & 0.8660 \\
%
%145 & LRA.03 old & Prob. with size & 8 & $\mathrm{XYCRy}$ & 24 & 0.2088 & 0.9764 & 0.9772 & 0.3297 & 0.9392 & 0.9455 & 0.8225 & 0.9267 & 0.8715 \\
%
145 & LRA.03 & Prob. with size & 8 & $\mathrm{XYCRy}$ & 24 & 0.2077 & 0.9765 & 0.9773 & 0.3296 & 0.9394 & 0.9456 & 0.8129 & 0.9267 & 0.8660 \\
%
%145 & LRA.04 old & Prob. with size & 8 & $\mathrm{XYCRy}\times 2$ & 48 & 0.2092 & 0.9762 & 0.9770 & 0.3273 & 0.9397 & 0.9460 & 0.8214 & 0.9200 & 0.8679 \\
%
145 & LRA.04 & Prob. with size & 8 & $\mathrm{XYCRy}\times 2$ & 48 & 0.2078 & 0.9765 & 0.9774 & 0.3299 & 0.9396 & 0.9458 & 0.8176 & 0.9267 & 0.8688 \\
%
%\midrule
%
%145 & \textcolor{red}{VQC-TEO} & \textcolor{red}{Train. Encoding \& Observable} & 8 & $\textcolor{red}{\mathrm{YZ}_{\mathrm{TE}}+\mathrm{CRy}}$ & 77 & \multicolumn{3}{c}{\textcolor{red}{second proposal by ChatGPT}} & 0.3947 & 0.9056 & \textcolor{red}{0.8906} & 0.8418 & 0.8867 & 0.8636 \\
%
%145 & \textcolor{red}{VQC-TO} & Prob. with size & 8 & $\mathrm{YCRy}\times 1$ & 53 & 0.5398 & 0.8504 & 0.8309 & 0.6004 & 0.7464 & 0.6955 & 0.6977 & 0.8000 & 0.7453 \\
%
%145 & \textcolor{red}{LRA-VQC-TO} & \textcolor{red}{Training Observable} & 8 & $\mathrm{YCRy}\times 1$ & 53 & 0.2099 & 0.9762 & 0.9770 & 0.3266 & 0.9392 & 0.9454 & 0.8214 & 0.9200 & 0.8679 \\
%
%145 & \textcolor{red}{LRA-VQC-TO} & \textcolor{red}{Training Observable} & 8 & $\mathrm{XYCRy}\times 2$ & 85 & 0.2093 & 0.9761 & 0.9769 & 0.3291 & 0.9387 & 0.9448 & 0.8176 & 0.9267 & 0.8688 \\
%
\midrule
10 & XGBoost & --- & --- & --- & --- & 0.1077 & 0.9976 & 0.9977 & 0.2156 & 0.9749 & 0.9743 & 0.8854 & 0.9267 & 0.9055 \\
\midrule
%
%10 & LRA.05 old & Cos-angle & 10 & $\mathrm{YCRy}$ & 20 & 0.3194 & 0.9339 & 0.9365 & 0.4195 & 0.8899 & 0.8859 & 0.7879 & 0.8667 & 0.8254 \\
%
10 & \textbf{LRA.05} & \textbf{Cos-angle} & 10 & $\bm{\mathrm{YCRy}}$ & 20 & 0.3194 & 0.9339 & 0.9371 & 0.4201 & 0.8899 & 0.8864 & 0.7857 & 0.8800 & 0.8302 \\
%
%10 & LRA.06 old & Cos-angle & 10 & $\mathrm{YCRy}\times 2$ & 40 & 0.3180 & 0.9343 & 0.9396 & 0.4159 & 0.8918 & 0.8914 & 0.7882 & 0.8933 & 0.8375 \\
%
10 & LRA.06 & Cos-angle & 10 & $\mathrm{YCRy}\times 2$ & 40 & 0.3114 & 0.9373 & 0.9416 & 0.4055 & 0.8972 & 0.8947 & 0.7976 & 0.8933 & 0.8428 \\
%
%10 & LRA.07 old & Cos-angle & 10 & $\mathrm{XYCRy}$ & 30 & 0.3205 & 0.9335 & 0.9363 & 0.4226 & 0.8880 & 0.8840 & 0.7824 & 0.8867 & 0.8313 \\
%
10 & LRA.07 & Cos-angle & 10 & $\mathrm{XYCRy}$ & 30 & 0.3192 & 0.9338 & 0.9374 & 0.4207 & 0.8892 & 0.8861 & 0.7857 & 0.8800 & 0.8302 \\
%
%10 & LRA.08 old & Cos-angle & 10 & $\mathrm{XYCRy}\times 2$ & 60 & 0.3213 & 0.9336 & 0.9362 & 0.4226 & 0.8885 & 0.8846 & 0.7929 & 0.8933 & 0.8401 \\
%
10 & LRA.08 & Cos-angle & 10 & $\mathrm{XYCRy}\times 2$ & 60 & 0.3108 & 0.9374 & 0.9416 & 0.4058 & 0.8970 & 0.8946 & 0.7917 & 0.8867 & 0.8365 \\
%
%10 & LRA.09 old & Prob. with size & 4 & $\mathrm{YCRy}$ & 8 & 0.3425 & 0.9264 & 0.9223 & 0.4477 & 0.8775 & 0.8692 & 0.7812 & 0.8333 & 0.8065 \\
%
10 & LRA.09 & Prob. with size & 4 & $\mathrm{YCRy}$ & 8 & 0.3427 & 0.9262 & 0.9215 & 0.4467 & 0.8776 & 0.8670 & 0.7812 & 0.8333 & 0.8065 \\
%
%10 & LRA.10 old & Prob. with size & 4 & $\mathrm{YCRy}\times 2$ & 16 & 0.3371 & 0.9289 & 0.9260 & 0.4411 & 0.8808 & 0.8722 & 0.7875 & 0.8400 & 0.8129 \\
%
10 & LRA.10 & Prob. with size & 4 & $\mathrm{YCRy}\times 2$ & 16 & 0.3380 & 0.9284 & 0.9240 & 0.4434 & 0.8801 & 0.8705 & 0.7875 & 0.8400 & 0.8129 \\
%
%10 & LRA.11 old & Prob. with size & 4 & $\mathrm{XYCRy}$ & 12 & 0.3413 & 0.9271 & 0.9217 & 0.4463 & 0.8774 & 0.8661 & 0.7862 & 0.8333 & 0.8091 \\
%
10 & LRA.11 & Prob. with size & 4 & $\mathrm{XYCRy}$ & 12 & 0.3427 & 0.9261 & 0.9213 & 0.4469 & 0.8776 & 0.8670 & 0.7812 & 0.8333 & 0.8065 \\
%
%10 & LRA.12 old & Prob. with size & 4 & $\mathrm{XYCRy}\times 2$ & 24 & 0.3355 & 0.9299 & 0.9240 & 0.4387 & 0.8823 & 0.8695 & 0.7862 & 0.8333 & 0.8091 \\
%
10 & LRA.12 & Prob. with size & 4 & $\mathrm{XYCRy}\times 2$ & 24 & 0.3381 & 0.9284 & 0.9235 & 0.4433 & 0.8800 & 0.8698 & 0.7862 & 0.8333 & 0.8091 \\
\bottomrule
\multicolumn{15}{@{}p{\linewidth}@{}}{%
    \footnotesize
    \textit{Note:} A compact notation is adopted for the ansatz circuits, which is the same used in Table~\ref{tab:new-VQC-table}.
}
\end{tabularx}
\end{table*}
%TTTTTTTTTTTTTTTTTTTTTTTTTTTTTTTTTTTTTTTTTT

%TTTTTTTTTTTTTTTTTTTTTTTTTTTTTTTTTTTTTTTTTT
\begin{table*}[!t]
\caption{Predictive performance of the classical and proposed quantum classification models on the final held-out test set. The model configurations selected on the validation set, highlighted in boldface in the preceding tables, are evaluated here without further model selection or hyperparameter adjustment. As discussed in Section~\ref{sec:egnome}, $N_{\mathrm{test}}=5{,}799$ samples were assigned to the test set. Performance metrics and optimization-convergence criterion (where applicable) are the same as those described in the caption of Table~\ref{tab:new-VQC-table}.}
\label{tab:new-test-table}

\centering
\scriptsize
\setlength{\tabcolsep}{2.5pt}
\renewcommand{\arraystretch}{1.15}

\begin{tabularx}{\textwidth}{
@{}rllrll
*{6}{X}
@{}
}
\toprule
\multicolumn{6}{c}{}
&
\multicolumn{6}{c}{Test}
\\
\cmidrule(lr){7-12}

$N_{\mathrm{F}}$
& Model
& Encoding
& $n$
& Ansatz
& $N_{\bm{\theta}}$
& BCE
& \shortstack{ROC--\\AUC}
& \shortstack{PR--\\AUC}
& P
& R
& \shortstack{$F_1$-score}
\\
\midrule

145 & XGBoost & --- & --- & --- & --- & 0.1676 & 0.9838 & 0.9765 & 0.8971 & 0.9692 & 0.9318 \\
\midrule
145 & \textbf{VQC.02} & \textbf{Amplitude} & 8 & $\bm{\mathrm{YCX}\times 4}$ & 32 & 0.5830 & 0.9007 & 0.8735 & 0.7583 & 0.8893 & 0.8186 \\
145 & \textbf{PQF.03} & \textbf{Amplitude} & 8 & --- & 1789 ($\bm{K=3}$) & 0.2019 & 0.9735 & 0.9500 & 0.8994 & 0.9585 & 0.9280 \\
%
%145 & PQF.xx & Amplitude & 8 & --- & 7459 ($K=4$) & 0.1943 & 0.9751 & 0.9552 & 0.9084 & 0.9652 & 0.9360 \\
%
%145 & \textbf{LRA.01 old} & \textbf{Prob. with size} & 8 & $\bm{\mathrm{YCRy}}$ & 16 & 0.2582 & 0.9611 & 0.9424 & 0.8640 & 0.9371 & 0.8991 \\
%
145 & \textbf{LRA.01} & \textbf{Prob. with size} & 8 & $\bm{\mathrm{YCRy}}$ & 16 & 0.2583 & 0.9611 & 0.9423 & 0.8628 & 0.9371 & 0.8984 \\
\midrule
10 & XGBoost & --- & --- & --- & --- & 0.1781 & 0.9846 & 0.9796 & 0.8893 & 0.9565 & 0.9217 \\
\midrule
10 & \textbf{VQC.22} & \textbf{Dense angle} & 5 & $\bm{\mathrm{XYCX}\times 4}$ & 40 & 0.3536 & 0.9326 & 0.9115 & 0.8269 & 0.9312 & 0.8760 \\
10 & \textbf{PQF.04} & \textbf{Angle} & 10 & --- & 31 ($\bm{K=1}$) & 0.1644 & 0.9867 & 0.9818 & 0.9006 & 0.9672 & 0.9327 \\
10 & \textbf{LRA.05} & \textbf{Cos-angle} & 10 & $\bm{\mathrm{YCRy}}$ & 20 & 0.3292 & 0.9333 & 0.9048 & 0.8363 & 0.9213 & 0.8768 \\
\bottomrule
\multicolumn{12}{@{}p{\linewidth}@{}}{%
    \footnotesize
    \textit{Note:} A compact notation is adopted for the ansatz circuits, which is the same used in Table~\ref{tab:new-VQC-table}.
}
\end{tabularx}
\end{table*}
%TTTTTTTTTTTTTTTTTTTTTTTTTTTTTTTTTTTTTTTTTT

\section{Numerical comparison of the classification models}
\label{sec:numerical-comparison}

\subsection{Ideal (statevector) simulations}
\label{sec:results-statevector}

The purpose of the present numerical experiment is to provide an initial, non-exhaustive comparison of the predictive performance of the classical and quantum classification models considered in this work. The analysis is not intended to establish a definitive ranking among the different approaches, which would require a broader investigation of architectures, hyperparameters, data partitions, and optimization strategies. Rather, it is designed to illustrate their relative behavior under a common and controlled computational setting. The reported results should therefore be interpreted as an illustrative benchmark of the considered models rather than as a
comprehensive optimization study. A more systematic assessment would require, among other aspects, repeated data partitions, cross-validation, statistical uncertainty estimates, and a dedicated hyperparameter search for each classical and quantum architecture.

As discussed in Section~\ref{sec:egnome}, the starting point of the present analysis is the thermoelectric-materials dataset developed within the Energy-GNoME project~\cite{DEANGELIS2025100605}, but we then intentionally restricted it to a training set with $N_{\mathrm{train}}=700$ samples and a validation set with $N_{\mathrm{valid}}=300$ samples. To obtain a genuinely held-out material-wise test set, we developed a test set which contains $N_{\mathrm{test}}=5{,}799$ samples associated with 1,026 previously unseen materials. The original samples are represented by $N_{\mathrm{F}}=145$ features. To facilitate the implementation of the proposed models on actual quantum hardware, SHAP (SHapley Additive exPlanations) \cite{Lundberg2017} was employed to select the $N_{\mathrm{F}}'=10<N_{\mathrm{F}}$ most informative features. The same data partition for both feature representations was used for all the models considered, namely XGBoost, the projected quantum feature model (PQF) described in Section~\ref{sec:projected}, the conventional variational quantum classifier (VQC) described in Section~\ref{sec:VQC}, and the logistic-regression-anchored variational quantum classifier (LRA) described in Section~\ref{sec:LRA-VQC}. All model configurations were specified before the final comparison, and the test set was not accessed during training, validation, or preprocessing. To prevent information leakage, all preprocessing parameters, including those associated with standardization and min--max scaling, were determined exclusively from the training set. The resulting transformations were subsequently applied to the validation and test samples without refitting. Likewise, all trainable classical and quantum parameters were optimized exclusively using the training data. During VQC and LRA-VQC optimization, training was terminated when the change in BCE satisfied $|\Delta\mathrm{BCE}|<10^{-5}$ for three consecutive checks, performed every ten objective-function evaluations.

The models were first evaluated on the training set through the binary cross-entropy (BCE), the area under the receiver operating characteristic curve (ROC--AUC), and the area under the precision--recall curve (PR--AUC). These quantities provide an indication of the quality of the optimization and, through comparison with the corresponding validation results, of the possible occurrence of overfitting. 
The predictive performance in the next step was assessed on the previously unseen validation set using BCE, ROC--AUC, PR--AUC, precision, recall, and the $F_1$-score. 
Unless otherwise specified, the discrete class predictions used to compute precision, recall, and the $F_1$-score were obtained by applying the common threshold $p(y=1\mid\mathbf{x})\geq0.5$. ROC--AUC measures the overall ability of a model to rank positive samples above negative ones independently of a particular classification threshold. 
PR--AUC is particularly informative for the present materials-screening problem when high-performing thermoelectric materials constitute a minority class. In this context, recall quantifies the fraction of genuinely promising materials identified by the model, whereas precision quantifies the fraction of the selected candidates that are actually promising. Thus, recall is related to the risk of overlooking potentially valuable materials, while precision reflects the computational or experimental effort spent on false-positive candidates.

All computations were performed using Python~3.13.18 within the Jupyter Notebook environment~\cite{Kluyver2016}. Data preprocessing and classical machine-learning models were implemented using scikit-learn~1.9.1~\cite{Pedregosa2011}. Quantum circuits and simulations were developed using Qiskit (Quantum Information Software Kit)~2.4.1~\cite{javadi-abhari_quantum_2024}, together with qiskit-algorithms~0.4.0 and qiskit-machine-learning~0.9.1.

Despite the logical sequence outlined in Section~\ref{sec:overarching}, we present first the results obtained with the conventional quantum classifier, followed by those achieved with the proposed quantum-inspired and quantum classifiers. The predictive performance on the training and validation sets is reported in Table~\ref{tab:new-VQC-table} for the VQC models, Table~\ref{tab:new-PQF-table} for the PQF models, and Table~\ref{tab:new-LRA-table} for the LRA-VQC models, which we call for simplicity LRA in the following. The consistently high performance achieved on the training sets indicates that the models are sufficiently expressive to capture the relevant relationships in the data, with no evident underfitting. At the same time, the limited degradation observed from training to validation indicates that the models retain good generalization capability and show no evidence of pronounced overfitting.

Among the available performance metrics, PR--AUC on the validation set was adopted as the primary criterion for model selection. By focusing explicitly on the positive class of interest, PR--AUC is especially informative for materials-screening applications, in which the objective is to identify as many promising candidates as possible while maintaining sufficiently high selection reliability. More specifically, within each model family and for each value of $N_{\mathrm{F}}$, we first retained all configurations whose validation PR--AUC was within $1\%$ of the maximum value, namely those satisfying
\begin{equation}
\mathrm{PR\text{-}AUC}
>
0.99\,\mathrm{PR\text{-}AUC}_{\max}.
\end{equation}
Among the retained configurations, the model with the smallest number of trainable parameters $N_{\bm{\theta}}$ was selected for evaluation on the final test set. This criterion favors a more parsimonious model whenever the improvement provided by a more complex architecture is marginal. The configurations selected through this validation procedure are highlighted in boldface in the preceding tables. Only these preselected models were subsequently evaluated on the final held-out test set, whose results are reported in Table~\ref{tab:new-test-table}.

The results obtained on the final held-out test set, reported in Table~\ref{tab:new-test-table}, can be summarized as follows. Consistently with the model-selection criterion discussed above, the following comparison is primarily based on PR--AUC, which is particularly informative for the present materials-screening application because it quantifies the trade-off between retrieving promising thermoelectric candidates and limiting false positives. For $N_{\mathrm{F}}=145$, PQF $\gtrsim$ LRA $>$ VQC, indicating that PQF slightly outperforms LRA, while LRA performs better than the conventional VQC. This result suggests that anchoring the quantum model to logistic regression is more effective than relying exclusively on the unconstrained optimization of the nonlinear VQC ansatz. PQF achieves the best PR--AUC among the investigated quantum approaches, although it remains slightly less effective than XGBoost applied directly to the original features. For $N_{\mathrm{F}}=10$, the ranking becomes PQF $>$ VQC $\gtrsim$ LRA, with VQC and LRA exhibiting substantially comparable PR--AUC values. Thus, even after SHAP-based feature reduction, PQF remains the most effective quantum approach and provides a PR--AUC only marginally better than that obtained by XGBoost on the original features.

Taken together, these findings indicate that, at least for the classification of thermoelectric materials considered here, the two potential benefits investigated in this work---namely, (i) the high dimensionality of quantum-state Hilbert spaces and (ii) the nonlinear feature transformations induced by quantum circuits---do not translate into a clear predictive advantage over a well-established classical classifier. One possible explanation is that the present classification task, although relevant to energy-materials discovery, may not possess sufficient structural complexity to benefit from the richer representations offered by quantum models and can already be effectively addressed by XGBoost. More generally, these results raise the question of the extent to which specifically quantum resources are advantageous for the classification of classical data. Quantum classifiers may be particularly beneficial when the data contain complex higher-order dependencies that are naturally represented by the adopted quantum feature map; however, such structures may not be ubiquitous in conventional classical datasets.

%FFFFFFFFFFFFFFFFFFFFFFFFFFFFFFFFFFFFFFFFFF
%\Figure[t](topskip=0pt, botskip=0pt, midskip=0pt)[width=\linewidth]
%{R-VQC-DA-YCX-star.png}
%{Variational Quantum Classifier (VQC) on real quantum hardware Lagrange. \textcolor{red}{[PIETRO: maybe @Matteo P reports the STAR topology here]}
%\label{fig:R-VQC-DA-YCX-star}}
%FFFFFFFFFFFFFFFFFFFFFFFFFFFFFFFFFFFFFFFFFF

%FFFFFFFFFFFFFFFFFFFFFFFFFFFFFFFFFFFFFFFFFF
%\Figure[t](topskip=0pt, botskip=0pt, midskip=0pt)[width=\linewidth]
%{R-VQC-DA-YCX-star-equiv.png}
%{Variational Quantum Classifier (VQC) on real quantum hardware Lagrange. \textcolor{red}{[PIETRO: maybe @Matteo P reports the STAR topology here]}
%\label{fig:R-VQC-DA-YCX-star-equiv}}
%FFFFFFFFFFFFFFFFFFFFFFFFFFFFFFFFFFFFFFFFFF

%FFFFFFFFFFFFFFFFFFFFFFFFFFFFFFFFFFFFFFFFFF
%\Figure[t](topskip=0pt, botskip=0pt, midskip=0pt)[width=0.999\linewidth]
%{R-VQC_ring.png}
%{Variational Quantum Classifier (VQC) on real quantum hardware Lagrange. \textcolor{red}{[PIETRO: first circuit executed on real hardware]}
%\label{fig:r-vqc-ring}}
%FFFFFFFFFFFFFFFFFFFFFFFFFFFFFFFFFFFFFFFFFF

%FFFFFFFFFFFFFFFFFFFFFFFFFFFFFFFFFFFFFFFFFF
%\Figure[t](topskip=0pt, botskip=0pt, midskip=0pt)[width=0.8\linewidth]
%{R-VQC_star.png}
%{Variational Quantum Classifier (VQC) on real quantum hardware Lagrange. \textcolor{red}{[PIETRO: second circuit executed on real hardware]}
%\label{fig:r-vqc-star}}
%FFFFFFFFFFFFFFFFFFFFFFFFFFFFFFFFFFFFFFFFFF

%FFFFFFFFFFFFFFFFFFFFFFFFFFFFFFFFFFFFFFFFFF
\begin{figure}[htbp]
    \centering
    \includegraphics[scale=1]{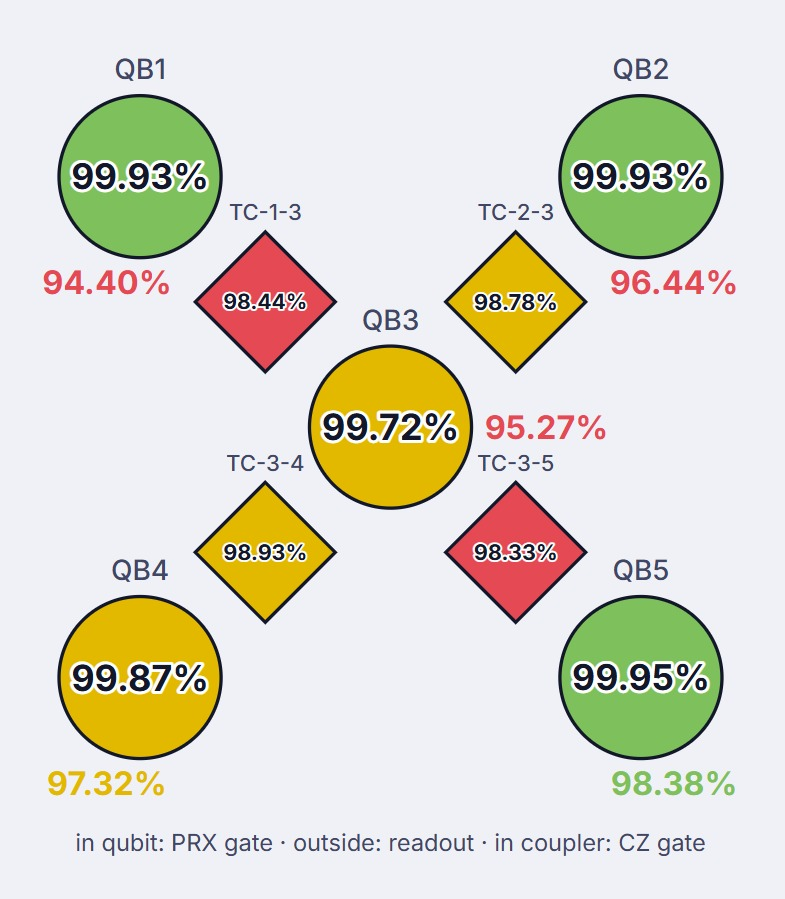}
    \caption{Calibration snapshot and native connectivity of the five-qubit Lagrange quantum processor. The percentages inside the circles denote the fidelities of the single-qubit phased-$R_x$ (PRX) gates, whereas those outside the circles indicate the corresponding readout fidelities. The diamonds represent the tunable couplers connecting the central qubit $\mathrm{QB3}$ to the peripheral qubits, with the corresponding controlled-$Z$ (CZ) gate fidelities reported inside. Diagnostic data acquired on November 20, 2026. Courtesy of the LINKS Foundation \cite{Viviani_LagrangeOperatingItalys_2026}.}
    \label{fig:2026-11-20_Lagrange}
\end{figure}
%FFFFFFFFFFFFFFFFFFFFFFFFFFFFFFFFFFFFFFFFFF

%FFFFFFFFFFFFFFFFFFFFFFFFFFFFFFFFFFFFFFFFFF
\begin{figure}[htbp]
    \centering
    \includegraphics[scale=0.50]{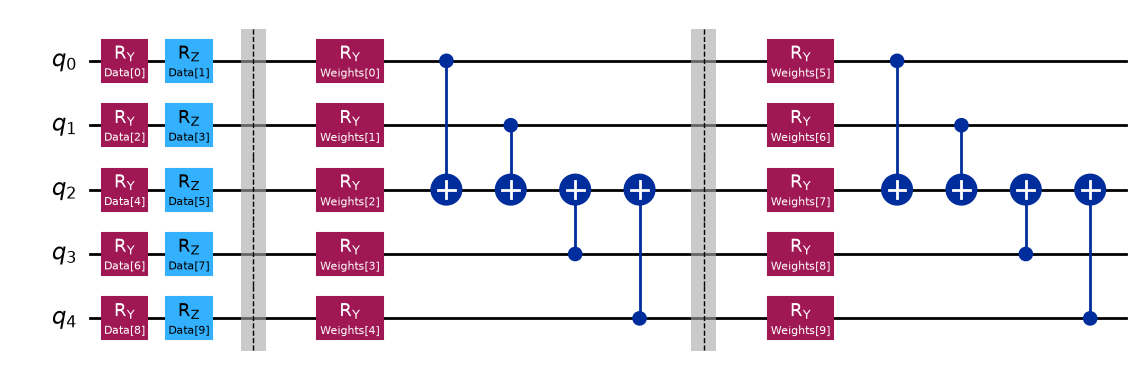}
    \caption{Example of an encoding \& trainable circuit for a Variational Quantum Classifier (VQC) running on hardware with star configuration. The feature map $U_{\varphi\phi}(\mathbf{x})$ implements dense-angle encoding, as described in Section~\ref{sec:encoding} and, in particular, in Eq.~(\ref{eqn:dense-angle-encoding}). The trainable circuit consists of two layers ($N_l=2$), each comprising a set of parameterized single-qubit $R_y$ rotations followed by controlled-$X$ gates arranged according to a star topology. In this example, $N_{\mathrm{F}}'=10$ and $N_{\bm{\theta}}=N_{\mathrm{A}}\times 2=10$.}
    \label{fig:VQC-DA-YCXstar_2L}
\end{figure}
%FFFFFFFFFFFFFFFFFFFFFFFFFFFFFFFFFFFFFFFFFF

%TTTTTTTTTTTTTTTTTTTTTTTTTTTTTTTTTTTTTTTTTT
\begin{table*}[!t]
\caption{Predictive performance of classification models on the training and test sets on real quantum hardware Lagrange. To facilitate the practical implementation, the datasets were randomly subsampled to obtain smaller datasets with $N_{\mathrm{train}}'=70$ and $N_{\mathrm{test}}'=30$. The performance metrics are the same as those described in the caption of Table~\ref{tab:new-VQC-table}. To reduce the computational cost, the models were selected through ideal statevector simulations limited to a fixed number of iterations, $N_{\mathrm{iter}}\sim 100$, since hardware noise prevents the reliable application of a smooth convergence criterion. For executions on the actual quantum hardware, each measurement circuit was sampled using $N_{\mathrm{shots}}=1024$ shots.}
\label{tab:new-Lagrange-table}

\centering
\scriptsize
\setlength{\tabcolsep}{2.5pt}
\renewcommand{\arraystretch}{1.15}

\begin{tabularx}{\textwidth}{
@{}rllrll
*{9}{X}
@{}
}
\toprule
\multicolumn{6}{c}{}
&
\multicolumn{3}{c}{Training}
&
\multicolumn{6}{c}{Ideal validation / Real test}
\\
\cmidrule(lr){7-9}
\cmidrule(lr){10-15}

$N_{\mathrm{F}}$
& Model
& Encoding
& $n$
& \shortstack{Ansatz}
& $N_{\bm{\theta}}$
& BCE
& \shortstack{ROC--\\AUC}
& \shortstack{PR--\\AUC}
& BCE
& \shortstack{ROC--\\AUC}
& \shortstack{PR--\\AUC}
& P
& R
& \shortstack{$F_1$-\\score}
\\
\midrule
10 & S-VQC.01 & Dense angle & 5 & $\mathrm{YCX}$ & 5 & 0.4878 & 0.9012 & 0.9216 & 0.5720 & 0.7822 & 0.8267 & 0.6471 & 0.7333 & 0.6875 \\
10 & S-VQC.02 & Dense angle & 5 & $\mathrm{YCX}\times 2$ & 10 & 0.4870 & 0.9037 & 0.9233 & 0.5708 & 0.8044 & 0.8336 & 0.6842 & 0.8667 & 0.7647 \\
10 & S-VQC.03 & Dense angle & 5 & $\mathrm{XYCX}$ & 10 & 0.4648 & 0.9535 & 0.9592 & 0.5386 & 0.8800 & 0.9138 & 0.6842 & 0.8667 & 0.7647 \\
10 & S-VQC.04 & Dense angle & 5 & $\mathrm{YCX}^\star$ & 5 & 0.6478 & 0.6571 & 0.7407 & 0.6803 & 0.5200 & 0.6064 & 0.5385 & 0.4667 & 0.5000 \\
10 & S-VQC.05 & Dense angle & 5 & $\mathrm{YCX}^\star\times 2$ & 10 & 0.4945 & 0.8824 & 0.8614 & 0.5266 & 0.8578 & 0.8494 & 0.6364 & 0.9333 & 0.7568 \\
\midrule
%
%10 & R-VQC.xx & Dense angle & 5 & $\mathrm{YCX}$ & 5 & \\
%
10 & R-VQC.01 & Dense angle & 5 & $\mathrm{XYCX}$ & 10 & 0.5986 & 0.8751 & 0.8415 & 0.6414 & 0.7778 & 0.7952 & 0.8571 & 0.4000 & 0.5455 \\
10 & R-VQC.02 & Dense angle & 5 & $\mathrm{YCX}^\star\times 2$ & 10 & 0.6691 & 0.6927 & 0.6704 & 0.6791 & 0.6578 & 0.6775 & 0.5652 & 0.8667 & 0.6842 \\
\midrule
%
%10 & S-PQF.01 & Dense angle & 5 & --- & 16 ($K=1$) & 0.121 & 1.000 & 1.000 & 0.211 & 0.991 & 0.992 & 1.000 & 0.867 & 0.929 \\
%
%10 & S-PQF.01 & Dense angle & 5 & processed by XGBoost & 16 ($K=1$) & 0.0930 & 1.0000 & 1.0000 & 0.6023 & 0.7778 & 0.7685 & 0.6316 & 0.8000 & 0.7059 \\
%
%10 & S-PQF.02 & Dense angle & 5 & processed by LR & 16 ($K=1$) & 0.2835 & 0.9502 & 0.9434 & 0.2965 & 0.9333 & 0.8709 & 0.8750 & 0.9333 & 0.9032 \\
%
%10 & S-PQF.03 & Dense angle & 5 & processed by XGBoost & 106 ($K=2$) & 0.0701 & 1.0000 & 1.0000 & 0.4285 & 0.8844 & 0.8747 & 0.9231 & 0.8000 & 0.8571 \\
%
%10 & S-PQF.04 & Dense angle & 5 & processed by LR & 106 ($K=2$) & 0.0924 & 1.0000 & 1.0000 & 0.2674 & 0.9467 & 0.9440 & 0.8824 & 1.0000 & 0.9375 \\
%
10 & S-PQF.01 & Dense angle & 5 & --- & 16 (1) & 0.0782 & 1.0000 & 1.0000 & 0.4052 & 0.9067 & 0.9232 & 0.7647 & 0.8667 & 0.8125 \\
10 & S-PQF.02 & Dense angle & 5 & --- & 106 (2) & 0.0542 & 1.0000 & 1.0000 & 0.5066 & 0.8756 & 0.8967 & 0.6500 & 0.8667 & 0.7429 \\
10 & S-PQF.03 (LR) & Dense angle & 5 & --- & 16 (1) & 0.2172 & 0.9722 & 0.9741 & 0.4358 & 0.8978 & 0.8949 & 0.7692 & 0.6667 & 0.7143 \\
10 & S-PQF.04 (LR) & Dense angle & 5 & --- & 106 (2) & 0.0739 & 1.0000 & 1.0000 & 0.7179 & 0.8356 & 0.8603 & 0.7222 & 0.8667 & 0.7879 \\
\midrule
%
%10 & R-PQF.01 & Dense angle & 5 & --- & 16 ($K=1$) & 0.117 & 1.000 & 1.000 & 0.221 & 0.991 & 0.992 & 0.933 & 0.933 & 0.933 \\
%
%10 & R-PQF.01 by XGBoost & Dense angle & 5 & --- & 16 ($K=1$) & 0.0950 & 1.0000 & 1.0000 & 0.5681 & 0.7733 & 0.7026 & 0.6667 & 0.8000 & 0.7273 \\
%
10 & R-PQF.01 & Dense angle & 5 & --- & 16 (1) & 0.0787 & 1.0000 & 1.0000 & 0.4635 & 0.8622 & 0.8772 & 0.8000 & 0.8000 & 0.8000 \\
\bottomrule
\multicolumn{15}{@{}p{\linewidth}@{}}{%
    \footnotesize
    \textit{Note:} The compact notation adopted for the ansatz circuits is the same as that used in Table~\ref{tab:new-VQC-table}. The superscript $^\star$ additionally denotes the star entanglement topology illustrated in Fig.~\ref{fig:VQC-DA-YCXstar_2L}. For all VQC circuits, the quantum score is evaluated using the observable $O'_2$, corresponding to a Pauli-$Z$ measurement on the central qubit $q=2$, namely $O'_2\equiv Z_2$.
}
\end{tabularx}
\end{table*}
%TTTTTTTTTTTTTTTTTTTTTTTTTTTTTTTTTTTTTTTTTT

\subsection{Results on real quantum hardware}
\label{sec:results-real}

Selected VQC and PQF configurations were also end-to-end  executed on Lagrange, an on-premises IQM Spark superconducting quantum computer jointly acquired by LINKS Foundation, Politecnico di Torino, and the Italian National Institute of Metrological Research (INRiM), and operated by LINKS Foundation~\cite{Viviani_LagrangeOperatingItalys_2026}. Lagrange comprises five superconducting transmon qubits arranged according to a star topology. The central qubit, denoted by $\mathrm{QB3}$, is connected through four independently tunable couplers to the peripheral qubits $\mathrm{QB1}$, $\mathrm{QB2}$, $\mathrm{QB4}$, and $\mathrm{QB5}$, whereas no direct coupling is available between pairs of peripheral qubits~\cite{Ronkko2024}. The QPU is operated at a temperature of approximately $\SI{30}{\milli\kelvin}$ and supports simultaneous frequency-multiplexed readout of the five qubits in the computational basis.

The native gate set consists of the single-qubit phased-$R_x$ gate, denoted by $\mathrm{PRX}$, and the two-qubit controlled-$Z$ gate~\cite{Ronkko2024,IQMSpark2026}. The former is defined as a rotation around an axis lying in the equatorial plane of the Bloch sphere (see Appendix \ref{sec:rotations}). Native $\mathrm{CZ}$ gates can be applied only between $\mathrm{QB3}$ and one of the four peripheral qubits. Consequently, interactions between two peripheral qubits require hardware-aware routing through the central qubit, potentially increasing the circuit depth and the accumulated gate error. Before execution, all VQC and PQF circuits were therefore transpiled into the native $\{\mathrm{PRX},\mathrm{CZ}\}$ gate set while accounting for the physical coupling map. Circuit submission and result retrieval were performed through the IQM software development kit and its Qiskit integration~\cite{IQMSDK2025}. Since gate and readout performance may vary between calibration cycles, the calibration data associated with the execution time should be considered when interpreting the hardware results. For reference, the manufacturer reports typical median fidelities of at least $99.9\%$ for single-qubit gates, $99.0\%$ for $\mathrm{CZ}$ gates, and $97\%$ for readout, with typical gate durations not exceeding $\SI{20}{\nano\second}$ and $\SI{60}{\nano\second}$ for single- and two-qubit gates, respectively~\cite{IQMSpark2026}.

A representative calibration snapshot of Lagrange is shown in Fig.~\ref{fig:2026-11-20_Lagrange}. The percentages reported inside the circles indicate the fidelities of the native single-qubit phased-$R_x$ (PRX) gates, whereas those outside the circles correspond to the readout fidelities. The diamond-shaped elements represent the tunable couplers, and the percentages reported inside them indicate the fidelities of the corresponding two-qubit controlled-$Z$ (CZ) gates. The single-qubit gate fidelities range from $99.72\%$ for $\mathrm{QB3}$ to $99.95\%$ for $\mathrm{QB5}$, with an average value of approximately $99.88\%$. The readout fidelities are appreciably lower and more heterogeneous, ranging from $94.40\%$ for $\mathrm{QB1}$ to $98.38\%$ for $\mathrm{QB5}$, with an average of approximately $96.36\%$. Finally, the CZ-gate fidelities range from $98.33\%$ for the $\mathrm{QB3}$--$\mathrm{QB5}$ connection to $98.93\%$ for the $\mathrm{QB3}$--$\mathrm{QB4}$ connection, with an average of approximately $98.62\%$. These values also highlight that the dominant device-level limitations arise from readout errors and, to a lesser extent, from two-qubit operations, whereas single-qubit gates exhibit consistently high fidelities. 

The connectivity map also shows that Lagrange has a native star topology, in which the central qubit $\mathrm{QB3}$ is directly coupled to each of the four peripheral qubits. Consequently, ansatz circuits based on star entanglement can be mapped more naturally onto the device, whereas ring-connected circuits require additional routing operations during transpilation. The values in Fig.~\ref{fig:2026-11-20_Lagrange} should nevertheless be regarded as a calibration snapshot rather than fixed device properties, since gate and readout fidelities may vary over time and between calibration cycles.

Coming back to the problem under consideration, to facilitate the practical implementation, the dataset effectively used for model training and testing on actual quantum hardware is
\begin{equation}
\label{Dused'}
\mathcal{D}'_{\mathrm{used}}
=
\mathcal{D}'_{\mathrm{train}}
\cup
\mathcal{D}'_{\mathrm{test}},
\end{equation}
with $N'_{\mathrm{used}}=70+30=100$ and with $N_{\mathrm{F}}'=10$ SHAP-reduced features (see Section~\ref{sec:egnome} for details). More specifically, the original training and validation sets, comprising $700$ and $300$ samples, respectively, were randomly subsampled to obtain $70$ training samples and $30$ test samples. The subsampling was performed independently within the two pre-existing, mutually exclusive partitions, thereby preserving their material-wise separation and preventing data leakage. In this hardware-oriented experiment, the $30$ held-out samples are used directly as the test set, without an intermediate validation stage, and remain completely unseen during model training. The feature map $U_{\varphi\phi}(\mathbf{x})$ was used to realize dense-angle encoding of the $N_{\mathrm{F}}'=10$ features, as described in Section~\ref{sec:encoding} and, in particular, in Eq.~(\ref{eqn:dense-angle-encoding}).

The native star connectivity of Lagrange motivates the introduction of a VQC ansatz with the same topology, as illustrated in Fig.~\ref{fig:VQC-DA-YCXstar_2L}. In particular, the central qubit $q=2$ is chosen as the common target of the controlled-$X$ gates, while the four peripheral qubits act as controls. This entangling pattern can be implemented directly using the native $\mathrm{CZ}$ gates of Lagrange by applying a Hadamard gate to the central qubit before and after the $\mathrm{CZ}$-star block. Indeed, for each peripheral qubit $q\in\{0,1,3,4\}$,
\begin{equation}
\label{eqn:CX-from-CZ-star}
H^{(2)}
\mathrm{CZ}^{(q,2)}
H^{(2)}
=
\mathrm{CX}^{(q,2)},
\end{equation}
where qubit $q$ is the control and qubit $2$ is the target. Since $H^{(2)}H^{(2)}=I^{\otimes 5}$, the same identity extends to the complete entangling block:
\begin{equation}
\label{eqn:CX-CZ-star-equivalence}
H^{(2)}
\left[
\prod_{q\in\{0,1,3,4\}}
\mathrm{CZ}^{(q,2)}
\right]
H^{(2)}
=
\prod_{q\in\{0,1,3,4\}}
\mathrm{CX}^{(q,2)}.
\end{equation}
Thus, the $\mathrm{CX}$-star ansatz is naturally compatible with the hardware connectivity and native two-qubit gate set, while the additional Hadamard operations can be decomposed into native single-qubit $\mathrm{PRX}$ rotations during transpilation. Consistently with this architecture, the readout is performed on the central target qubit using
\begin{equation}
\label{eqn:central-star-observable}
O'_2
=
I^{\otimes 2}\otimes Z\otimes I^{\otimes 2}
\equiv Z_2.
\end{equation}
The central qubit is thereby used to collect the correlations generated by its interactions with all four peripheral qubits.

Considering the VQC results reported in Table~\ref{tab:new-VQC-table}, and excluding the trainable $\mathrm{YCRy}$ ansatz because it produces excessively deep circuits after hardware-aware transpilation, the most promising architectures are based on the $\mathrm{YCX}$ and $\mathrm{XYCX}$ ansatz circuits. Moreover, the native star connectivity of Lagrange motivates the investigation of the corresponding hardware-oriented configurations, in which the $\mathrm{CX}$ entangling gates are arranged according to the same topology. These configurations are denoted by $\mathrm{YCX}^{\star}$, as illustrated in Fig.~\ref{fig:VQC-DA-YCXstar_2L}, and $\mathrm{XYCX}^{\star}$. To limit the computational cost, these models were evaluated through ideal statevector simulations using a fixed optimization budget of approximately $N_{\mathrm{iter}}\sim100$ iterations. A fixed iteration budget was adopted because hardware noise makes the subsequent application of a smooth convergence criterion to real-QPU executions unreliable. The configurations satisfying these requirements are identified by the prefix ``S-'' in Table~\ref{tab:new-Lagrange-table}. Among them, the highest validation PR--AUC values are obtained using the $\mathrm{XYCX}$ and $\mathrm{YCX}^{\star}\times2$ trainable circuits. Both configurations contain $N_{\bm{\theta}}=10$ trainable parameters.

Turning to the PQF approach, we considered configurations whose measurement cost is comparable to or lower than that of the VQC executions, which require approximately $100\times70$ circuit evaluations. The most suitable PQF configurations for execution on Lagrange are therefore those with maximum Pauli weights $K=1$ and $K=2$. For $n=5$ qubits, the $K=1$ Pauli expansion contains $N_1=3n=15$ single-qubit observables, corresponding to one $X$, $Y$, and $Z$ operator for each qubit, and thus requires approximately $70\times15$ expectation-value estimations over the training set. The additional parameter included in the reported value $N_{\bm{\theta}}=N_1+1=16$ is the classical bias. Hence, $16$ represents the total number of classifier parameters, whereas only $15$ quantum observables must be estimated on the hardware. For $K=2$, the expansion contains $N_2=3n\,(3n-1)/2=105$ nontrivial Pauli observables, corresponding to approximately $70\times105$ expectation-value estimations, while the inclusion of the classical bias yields $N_{\bm{\theta}}=N_2+1=106$. As reported in Table~\ref{tab:new-Lagrange-table}, the same PQF configurations were also assessed through statevector simulations using either logistic regression (LR) or XGBoost as the final classical post-processor. Among these simulations, denoted by the prefix ``S-'', the highest validation PR--AUC is achieved by the $K=1$ configuration employing standard XGBoost, with $N_{\bm{\theta}}=16$.

The LRA-VQC approach was excluded because it is not currently formulated for dense-angle encoding. Its alternative implementations would require either $10$ qubits or a four-qubit probability-with-size encoding based on a comparatively deep state-preparation circuit, both of which are impractical for the present hardware experiment. 

The most promising VQC and PQF configurations identified through the preceding analysis constitute the set of models selected for execution on Lagrange and are denoted by the prefix ``R-'' in Table~\ref{tab:new-Lagrange-table}. For executions on the actual quantum hardware, each measurement circuit was sampled using $N_{\mathrm{shots}}=1024$ shots. As expected, hardware noise generally reduces the PR--AUC relative to the corresponding ideal statevector results. Moreover, although the star topology can be transpiled more efficiently onto the native hardware connectivity, it yields a lower PR--AUC than the ring topology in the experiments considered here. Conversely, the star topology performs more favorably in terms of the $F_1$-score, whereas the opposite trend is observed for precision, P. These comparisons should nevertheless be regarded as qualitative rather than conclusive, since a statistically robust assessment would require multiple independent executions. Such an analysis is beyond the available computational budget, as each complete experiment requires approximately one hour of QPU time $t_{\mathrm{QPU}}$ (corresponding, at current rates, to a cost ranging from approximately one to several thousand euros).

To verify that the QPU time required by the present experiments is consistent with that reported for comparable executions on the same hardware, we focus on training, which is the most computationally demanding stage, and neglect the final evaluation on the test set. For each R-VQC model listed in Table~\ref{tab:new-Lagrange-table}, we estimate the total number of circuits evaluated during training as $N_{\mathrm{circuit}}'\sim N_{\mathrm{iter}}\times N_{\mathrm{train}}'=7{,}000$. Since each circuit must be measured repeatedly over $N_{\mathrm{shots}}$ shots, the average effective QPU time per circuit shot can be estimated as $t_{\mathrm{QPU}}/(N_{\mathrm{circuit}}'\times N_{\mathrm{shots}})\sim\SI{0.5}{\milli\second}$. This estimate includes the hardware operations required to initialize and measure the qubits, which typically account for most of the execution time. Despite its approximate nature, the resulting value is consistent with the execution times reported in Ref.~\cite{Sestini2026}, where a more detailed analysis of QPU timing is provided.

\section{Conclusions}
\label{sec:conclusions}

In this work, we investigated quantum approaches to the binary classification of thermoelectric materials, considering both their methodological foundations and their end-to-end implementation. In addition to the conventional variational quantum classifier (VQC), two alternative hybrid quantum--classical strategies were examined. The first is the Projected Quantum Features (PQF) method, a simplified projected-feature approach in which expectation values measured on an encoded quantum state are directly employed as transformed features for a subsequent classical classifier. The second, and most distinctive, contribution is the logistic-regression-anchored variational quantum classifier (LRA-VQC). Rather than replacing an established classical model, LRA-VQC uses logistic regression as a baseline and trains a variational quantum circuit to describe possible nonlinear corrections to its predictions. This construction provides a controlled route for introducing quantum-generated nonlinearities only when they are beneficial. In the numerical experiments considered here, LRA-VQC frequently outperformed the corresponding unconstrained VQC, although its advantage depended on the encoding and circuit architecture. More generally, the anchoring principle is not specific to thermoelectric materials and could be applied to other classification problems for which a reliable classical baseline is available.

A further contribution of this work is the formulation of VQC, PQF, and LRA-VQC within a common theoretical framework. The three approaches can be distinguished according to where the trainable parameters are assigned: to the quantum state through a variational circuit, to the observable or classical readout acting on quantum-generated features, or jointly to both components. This unified perspective clarifies the relationships among apparently different quantum-classification strategies and provides a systematic basis for designing further hybrid models. In particular, it emphasizes that the relevant modeling choice is not merely the structure of the quantum circuit, but the allocation and interpretation of the learnable degrees of freedom throughout the complete classification pipeline.

The ideal statevector results show that PQF is generally the most competitive of the quantum approaches considered for this application, while LRA-VQC can provide an effective and more structured alternative to a conventional VQC. Nevertheless, the results also confirm that well-established classical models, particularly XGBoost, remain extremely competitive for the present thermoelectric-materials dataset. Neither the high dimensionality of the quantum-state Hilbert space nor the nonlinear transformations generated by the investigated quantum circuits produced a systematic advantage over the strongest classical baseline. This outcome suggests that the practical value of quantum classifiers remains strongly dependent on the structure of the data, the adopted encoding, the accessible observables, and the compatibility between the circuit-induced feature map and the underlying classification problem.

Beyond ideal simulations, selected VQC and PQF configurations were executed on the Lagrange quantum computer. Importantly, the quantum hardware was used not only for inference with previously optimized parameters, but throughout the complete machine-learning workflow, including model training and the subsequent prediction of previously unseen samples. As expected, finite-shot statistics, hardware noise, and transpilation constraints reduced the predictive performance relative to ideal statevector simulations. The experiments nevertheless demonstrate the feasibility of carrying out a complete quantum-classification workflow for thermoelectric-materials screening on currently available quantum hardware. To the best of our knowledge, this is the first demonstration of a quantum algorithm for thermoelectric-materials identification in which all stages of the machine-learning workflow, including training and testing, are performed on real quantum hardware.

The hardware results should be interpreted as a proof of feasibility rather than as evidence of quantum advantage. The limited number of samples, the restricted optimization budget, and the cost of repeated QPU executions prevented an extensive statistical analysis over multiple independent runs. Future work should therefore investigate more noise-resilient circuit architectures, measurement-reduction strategies, trainable and repeated data encodings, error-mitigation techniques, and larger hardware experiments. Particularly promising directions include the application of the LRA principle to other classical baselines and scientific domains, as well as the development of projected-feature models that retain the favorable performance of PQF while reducing the number of required measurements.

%%%%%%%%%%%%%%%%%%%%%%%%%%%%%%%%%%%%%%%%%%%%%%%%%%%%%%
%%%%%%%%%%%%%%%%%%%%%%%%%%%%%%%%%%%%%%%%%%%%%%%%%%%%%%
\section*{Acknowledgment}

The authors acknowledge access to the IQM quantum computer Lagrange~\cite{Viviani_LagrangeOperatingItalys_2026}, hosted at Politecnico di Torino (Turin, Italy), managed by the LINKS Foundation and jointly acquired by Politecnico di Torino, the Fondazione LINKS, and Istituto Nazionale di Ricerca Metrologica (INRiM), which was used to perform the quantum experiments reported in this work. In particular, the authors gratefully acknowledge Paolo Viviani from the LINKS Foundation for providing valuable technical support.

%%%%%%%%%%%%%%%%%%%%%%%%%%%%%%%%%%%%%%%%%%%%%%%%%%%%%%
%%%%%%%%%%%%%%%%%%%%%%%%%%%%%%%%%%%%%%%%%%%%%%%%%%%%%%

\section*{Code availability}\label{sec:code-availability}
The complete workflow is implemented through Python scripts and Jupyter Notebooks, encompassing dataset assembly, preprocessing, model definition, training, and performance evaluation. To support transparency and reproducibility, the complete source code will be made publicly available in an online repository upon publication.

%%%%%%%%%%%%%%%%%%%%%%%%%%%%%%%%%%%%%%%%%%%%%%%%%%%%%%
%%%%%%%%%%%%%%%%%%%%%%%%%%%%%%%%%%%%%%%%%%%%%%%%%%%%%%

\section*{Declaration of Generative AI and AI-Assisted Technologies}

During the preparation of this manuscript, the authors used OpenAI's ChatGPT to support language editing, the drafting and revision of selected passages, and the development and debugging of portions of the computational code. All AI-assisted material, including technical statements, mathematical expressions, references, code, and numerical results, was critically reviewed, independently verified, and revised by the authors as appropriate. The authors take full responsibility for the accuracy, originality, and integrity of the final manuscript.

%%%%%%%%%%%%%%%%%%%%%%%%%%%%%%%%%%%%%%%%%%%%%%%%%%%%%%
%%%%%%%%%%%%%%%%%%%%%%%%%%%%%%%%%%%%%%%%%%%%%%%%%%%%%%

\section{Appendix}

\subsection{Rotations on the Bloch sphere}
\label{sec:rotations}

A generic pure single-qubit state can be represented as a point on the surface of the Bloch sphere~\cite{nielsen_quantum_2010}. More specifically, up to a physically irrelevant global phase, it can be
written as
\begin{equation}
\label{eqn:Bloch-state}
\ket{\psi}
=\cos\left(\frac{\varphi}{2}\right)\ket{0}
+e^{i\phi}
\sin\left(\frac{\varphi}{2}\right)\ket{1},
\end{equation}
where $0\leq\varphi\leq\pi$ and $0\leq\phi<2\pi$ are the polar and azimuthal angles, respectively.

Single-qubit operations can be described in terms of elementary rotations about the Cartesian axes of the Bloch sphere, denoted by $R_x(\vartheta)$, $R_y(\vartheta)$, and $R_z(\vartheta)$. Although the
rotation angle $\vartheta$ can take any real value, it may be restricted to a $2\pi$ interval owing to the periodicity of the corresponding physical transformation. These rotation gates are defined as
\begin{equation}
\label{Rx}
R_x(\vartheta)
=
\begin{bmatrix}
\cos(\vartheta/2) & -i\sin(\vartheta/2) \\
-i\sin(\vartheta/2) & \cos(\vartheta/2)
\end{bmatrix},
\end{equation}
\begin{equation}
\label{Ry}
R_y(\vartheta)
=
\begin{bmatrix}
\cos(\vartheta/2) & -\sin(\vartheta/2) \\
\sin(\vartheta/2) & \cos(\vartheta/2)
\end{bmatrix},
\end{equation}
and
\begin{equation}
\label{Rz}
R_z(\vartheta)
=
\begin{bmatrix}
\exp(-i\vartheta/2) & 0 \\
0 & \exp(i\vartheta/2)
\end{bmatrix}.
\end{equation}
All three definitions follow from the common expression
\begin{equation}
\label{eqn:Pauli-rotation}
R_\alpha(\vartheta)
=
\exp\left(
-i\frac{\vartheta}{2}\sigma_\alpha
\right)
=
\cos\left(\frac{\vartheta}{2}\right)I
-i\sin\left(\frac{\vartheta}{2}\right)\sigma_\alpha,
\end{equation}
where $\alpha\in\{x,y,z\}$  and $\sigma_\alpha\in\{X,Y,Z\}$ denotes the Pauli matrix associated with the
corresponding rotation axis. The second equality follows from the property $\sigma_\alpha^2=I$ and represents the matrix counterpart of Euler's formula. More formally, the anti-Hermitian operators $-i\sigma_\alpha/2$ generate single-qubit rotations and form a basis of the Lie algebra $\mathfrak{su}(2)$ associated with the group $SU(2)$~\cite{nielsen_quantum_2010}.

It is important to distinguish the unrestricted parameters of the rotation gates from the angular coordinates conventionally used to represent a state on the Bloch sphere. While the gate parameter $\vartheta$ may take any real value, a unique geometrical representation of a pure single-qubit state employs a polar angle $\varphi\in[0,\pi]$ and an azimuthal angle $\phi\in[0,2\pi)$. Accordingly, when $R_y$ is used to encode a non-periodic scalar feature as a polar angle, restricting its argument to $[0,\pi]$ provides a non-redundant mapping. By contrast, an $R_z$ rotation naturally encodes an azimuthal angle over $[0,2\pi)$, although the endpoints represent the same physical direction and may therefore introduce a wrap-around effect for non-periodic features. Moreover, $R_z$ can encode observable information only when acting on a superposition state: its action on $\ket{0}$ or $\ket{1}$ changes only the phase, and a subsequent measurement in the computational basis alone is insensitive to this phase.

%\newpage
%\bibliographystyle{unsrtnat}
%\bibliography{references}

\end{document}